\documentclass{article} 
\usepackage{nips13submit_e,times}

\usepackage{url}

\usepackage{amsmath}
\usepackage{graphicx}
\usepackage{caption}
\usepackage{subcaption}
\usepackage{url}
\usepackage{natbib}
\usepackage{hyperref}
\usepackage[noabbrev,capitalise]{cleveref}
\usepackage{lipsum}
\usepackage{natbib}
\usepackage{amssymb}
\usepackage{multicol}
\usepackage{bm}
\usepackage{wrapfig}
\usepackage{tikz}
\usepackage{float}
\usepackage{subcaption}
\usepackage{mathtools}
\usepackage{amsfonts}
\usepackage{bbm}

\usepackage{algorithmicx}
\usepackage[noend]{algpseudocode}
\usepackage{algorithm}

\algrenewcommand\alglinenumber[1]{{\sf\scriptsize{\color{brown} #1}}}
\algrenewcommand\algorithmicrequire{\textbf{Input:}}

\newtheorem{theorem}{Theorem}
\usepackage{xcolor}
\definecolor{forestgreen}{rgb}{0.13, 0.55, 0.13}

\usepackage{soul}

\usepackage{setspace}
\title{Discovering Active Subspaces for High-Dimensional Computer Models}

\author{
Kellin Rumsey$^1$ \\
Statistical Sciences\\
Los Alamos National Laboratories \\
Los Alamos, NM 87545 \\
\texttt{knrumsey@lanl.gov} \\
\And
Devin Francom \\
Statistical Sciences\\
Los Alamos National Laboratories \\
Los Alamos, NM 87545 \\
\texttt{dfrancom@lanl.gov} \\
\And
Scott Vander Wiel \\
Statistical Sciences\\
Los Alamos National Laboratories \\
Los Alamos, NM 87545 \\
\texttt{scottv@lanl.gov}
}

\usepackage[normalem]{ulem}
\nipsfinalcopy 

\begin{document}

\maketitle

\normalsize

\onehalfspacing
\vspace*{-30pt}

\begin{abstract}
\small
Dimension reduction techniques have long been an important topic in statistics, and active subspaces (AS) have received much attention this past decade in the computer experiments literature. The most common approach towards estimating the AS is to use Monte Carlo with numerical gradient evaluation. While sensible in some settings, this approach has obvious drawbacks. Recent research has demonstrated that active subspace calculations can be obtained in closed form, conditional on a Gaussian process (GP) surrogate, which can be limiting in high-dimensional settings for computational reasons. In this paper, we produce the relevant calculations for a more general case when the model of interest is a linear combination of tensor products. These general equations can be applied to the GP, recovering previous results as a special case, or applied to the models constructed by other regression techniques including multivariate adaptive regression splines (MARS). Using a MARS surrogate has many advantages including improved scaling, better estimation of active subspaces in high dimensions and the ability to handle a large number of prior distributions in closed form. In one real-world example, we obtain the active subspace of a radiation-transport code with 240 inputs and 9,372 model runs in under half an hour.
\normalsize
\end{abstract}

\doublespacing
\newpage

\standardsize

\section{Introduction}
Active Subspace Methods (ASM's) have become an essential part of the Uncertainty Quantification (UQ) toolkit in recent years. In the last decade alone, these methods have been applied to a large number of diverse domains including aerospace shape optimization, turbulent combustion, hydrological models, epidemiology, automotive design, lithium ion batteries, single-diode solar cells, turbomachinery, indicators for Alzheimers disease and cardiovascular anomalies \citep{lukaczyk2014active, ji2019quantifying, jefferson2015active, hielscher2018framework, othmer2016active, constantine2017time, constantine2015discovering,seshadri2018turbomachinery, batta2021uncovering, tezzele2018combined}.


Fundamentally, an active subspace is the space spanned by a set of orthogonal ``important direction" vectors in the input space for a function $f$. In particular, moving along these directions will cause the largest possible changes in the output of $f$. In the popular framework of \cite{constantine2015activebook}, these vectors are taken to be the leading eigenvectors of the expected outer product of the gradient of $f$. The corresponding active subspace can be used to (i) define global sensitivity metrics for the original inputs \citep{constantine2017global} (ii) circumvent the ``curse of dimensionality" by providing a low-dimensional approximation of a computer model \citep{vohra2019active, constantine2014activekriging} (iii) accelerate MCMC in a Bayesian analysis of a computer model \citep{constantine2016accelerating} (iv) improve the memory and time footprint of a surrogate model and (v) improve the design of computer model experiments \citep{navaneeth2022surrogate, wycoff2021sequential}.

Discovering the active subspace relies heavily on the gradients of the computer model (function) at hand. In many non-trivial applications, the computer model is treated as a compute-intensive ``black-box" and exact gradient calculations are unavailable. Thus there are two primary methods for attempted learning of the active subspace. The first is to approximate the gradients numerically (e.g., using finite differences or automatic differentiation) and to couple these with Monte Carlo to arrive at an estimate of the active subspace. There are several limitations to this approach, including:
\begin{itemize}
    \item For models with a large number of parameters, both the gradient approximations and the Monte Carlo procedure can be time consuming and possibly inaccurate. 
    \item Uncertainty is introduced into the estimate in both the approximation of the gradient and the Monte Carlo procedure. It can be difficult or expensive to account for these uncertainties. 
    \item The computer experiment must be specifically tailored to the input distribution for which the corresponding active subspace is desired. This can be challenging if the practitioner is interested in the active subspace corresponding to several priors, as in robust Bayes \citep{berger1990robust}, or if the practitioner hopes to design the computer experiment sequentially \citep{wycoff2021sequential}. In other cases, like when a black-box model is informed only through a ``legacy" dataset, incorporating the input distribution into the design requires remedies such as importance sampling, which is an inefficient use of computational resources and can limit the accuracy of the approximation.
\end{itemize}
For these reasons, it is common to replace the numerical derivative approximations with those from a computer model surrogate \citep{constantine2015activebook}. In most cases, this surrogate is used in tandem with the Monte Carlo approach, but there are some notable exceptions. For instance, \cite{xia2008multiple} develops a coherent model based approach based on single index models, which estimates the low dimensional structure directly. Several other authors have developed single-step procedures to learn ``influential directions" of some kind. Projection pursuit \citep{friedman1981projection} and sliced-inverse regression \citep{li1991sliced} are two well-known examples with similar goals, although neither directly uses the gradient of $f$.  The average derivative estimation of \cite{hardle1989investigating} is nearly the same as the ASM we refer to here, with the order of the outer product and expectation interchanged. Recently, \cite{navaneeth2022surrogate} incorporate sparse regression into active subspace identification. 

Our approach is most similar to the recent work by \cite{wycoff2021sequential}, in which a closed form is derived for the active subspace given a Gaussian process surrogate. This allows for accurate and efficient computation of the active subspace whenever a Gaussian process can be trained (also accurately and efficiently) on computer model output. This approach also allows for sequential design of the computer experiment and does not require a strict relationship between the input distribution (or prior) of interest and the computer experimental design. In this paper, we derive a more general closed form solution for the active subspace when the function of interest can be written (or approximated) as a linear combination of tensor products of generic functions used as building blocks. In many cases, this class of functions will be a universal approximator \citep{lin1992canonical} indicating that the solution can, in theory, be applied to any continuous function (with a compact domain) to obtain an arbitrarily good approximation of the active subspace. The Multivariate Adaptive Regression Splines (MARS) algorithm \citep{friedman1991} and the Bayesian MARS (BMARS) algorithm \citep{denison1998bayesian, francom2020bass} produce models which represent a special case of this function class.  In practice we can use MARS (or BMARS) as a surrogate to a computer model $f$, and obtain an estimate of the active subspace. The primary contributions of this work are given as follows. 
\begin{itemize}
    \item {\it Efficiency and Scalability}. The MARS surrogate scales excellently with the number of computer model runs and the number of input parameters. While Gaussian processes are sometimes considered the gold standard for surrogate modeling, they suffer from notoriously poor scaling. When the number of inputs is very large, a large amount of training data is required, and the cubic scaling of a GP can render it infeasible. In addition to the cost associated with training a GP, the GP-based active subspace estimation procedure proposed in \cite{wycoff2021sequential} scales poorly with the number of inputs (see \cref{fig:example1}).
    
    On the other hand, training a MARS model is nearly always fast, and can handle orders of magnitude more training data than a standard GP. Modern approaches to Bayesian MARS can also leverage parallel tempering and flexible likelihoods (see Sections 3.1-3.2 in the SM for examples) to improve predictive performance and posterior mixing \citep{francom2020bass, rumsey2023generalized}. In the context of Bayesian MARS, we also note that the full posterior distribution for the active subspace can be obtained at a sub-linear amortized cost per MCMC iteration (see Sections 3.3 and 6-8 of the SM). 
    
    \item {\it Flexibility.} Our method is flexible in terms of the probability measure specified for the input parameters. We provide a closed form solution when the prior distribution for the inputs is Uniform, Gaussian, $t$, Gamma or Beta. We also show how the multivariate Gaussian distribution can be handled, accounting for correlation between inputs. Additionally, we can handle a finite mixture of these distributions, with interval truncation if needed. This collection of priors represents a wide variety of important cases, such as constrained support and calibration-type problems where we can only access samples from the desired measure. Finally, we can discover the active subspace with respect to many different measures, without needing to re-train the surrogate model. 
    
    \item {\it Generality.} The results in this paper (especially \cref{sec:general}) are general and can be applied to a large number of surrogate functions. Although we emphasize the use of MARS as a surrogate, the results in \cref{sec:general} can be applied to a wide variety of emulators. In fact, the Gaussian process (with a separable kernel) can be written as a linear combination of tensor products \citep{higdon2002space, gramacy2020surrogates}, and the GP-based approach of \cite{wycoff2021sequential} can be recovered as a special case. 
\end{itemize}

The rest of this paper is organized as follows. In \cref{sec:background}, we review the relevant background related to Active Subspaces and MARS regression. In \cref{sec:methods}, we derive the closed form solution to the primary active subspace quantities in a general setting and for MARS. For MARS and BMARS, our formulation includes results for a large class of interesting prior distributions over the inputs. In Section 4 we demonstrate the functionality of our methods using several illustrative examples and two real data sets. Discussion and concluding remarks are given in Section 5. An R package providing an implementation of the methods discussed in \cref{sec:methods} can be found at \url{https://github.com/knrumsey/concordance}.

\section{Background}
\label{sec:background}
We begin by providing some brief background on active subspace methods and MARS. For a more detailed discussion refer to \cite{constantine2015activebook} for ASM's, \cite{friedman1991} for MARS and \cite{francom2020bass} for BMARS. 

\subsection{Active Subspaces}
\label{sec:active}
Approximation based on lower-dimensional projections is not a new idea. These ideas are foundational in established methods such as Projection Pursuit and Sliced Inverse Regression \citep{friedman1981projection, li1991sliced}. The goal of the active subspace can be intuitively understood as trying to find directions in the input space such that moving along these directions causes the function to change the most on average (with respect to some measure). Likewise, we will often find some inactive directions within which the function changes very little, and the hope is that these can be ignored without much consequence. More specifically, if $\bm x \in \mathbb R^p$ is a vector of inputs to the scalar valued function $f$, then the goal of ASM is to find a projection matrix $\bm W_1 \in \mathbb R^{p \times r}$ (with $r \leq p$) such that 
\begin{equation}
\label{eq:approx}
f(\bm x) \approx \tilde f(\bm W_1^\intercal \bm x)
\end{equation}
for some suitable $\tilde f: \mathbb R^r \rightarrow \mathbb R$. To formally define $\bm W_1$, first consider the matrix
\begin{equation}
\label{eq:C}
    \bm C = \mathbb E\left[\nabla f(\bm x) \nabla f(\bm x)^\intercal\right] = \int \nabla f(\bm x) \nabla f(\bm x)^\intercal \rho(\bm x) d\bm x.
\end{equation}
$\bm C$ is a $p \times p$ symmetric, positive semi-definite matrix, which can be read as the expected outer product of the gradient of $f(\bm x)$. The expectation is taken with respect to a measure $\rho(\bm x)$. This measure, often viewed as a prior distribution, is given little attention in the ASM literature. In this work, we will discuss $\rho(\bm x)$ thoroughly, and derive formulas for a wide variety of choices. 

The active subspace is now defined by the eigenvectors of $\bm C$ as
\begin{equation} \label{eq:eigendecomp}
    \bm C = \bm W\bm\Lambda \bm W^\intercal = 
    \begin{bmatrix}
    \bm W_1 & \bm W_2
    \end{bmatrix}
        \begin{bmatrix}
    \bm \Lambda_1 & \bm 0 \\
    \bm 0 & \bm\Lambda_2
    \end{bmatrix}
    \begin{bmatrix}
    \bm W_1^\intercal \\ \bm W_2^\intercal
    \end{bmatrix},
\end{equation}
where $\bm W = [\bm w_1, \ldots, \bm w_p]$ is an orthonormal matrix of eigenvectors and $\bm \Lambda = \text{Diag}(\lambda_1, \ldots, \lambda_p)$ is a diagonal matrix of eigenvalues such that $\lambda_i \geq \lambda_{i+1}$. Intuition can be gained by noting that the eigenpairs must satisfy
\begin{equation}
    \lambda_i = \bm w_i^\intercal \bm C\bm w_i = \int[\bm w_i^\intercal \nabla f(\bm x)]^2 \rho(\bm x)d\bm x.    
\end{equation}
This shows that $\lambda_i = 0$ if and only if $f(\bm x)$ is constant along the direction $\bm w_i$. In general, larger eigenvalues indicate larger changes on average in the function $f(\bm x)$ along the direction of the corresponding eigenvector.

The second equality of \cref{eq:eigendecomp} partitions $\bm W = [\bm W_1 \quad \bm W_2]$ with $\bm W_1 \in \mathbb R^{p \times r}$ and $\bm W_2 \in \mathbb R^{p \times (p-r)}$. If the eigenvalues in $\bm \Lambda _2$ are sufficiently small, then $\tilde f(\bm W_1^\intercal \bm x)$ should be a good low-dimensional approximation for $f(\bm x)$.

\subsection{Multivariate Adaptive Regression Splines}
\label{sec:mars}

Consider data pairs $({\bm x}_i, y_i)$, $i=1,2\ldots n$ with ${\bm x}_i \in \mathcal X \subset \mathbb R^p$ and $y_i = f({\bm x}_i)$ and assume, without loss of generality, that each input has been rescaled to the interval $[0, 1]$. MARS regression takes 
\begin{equation}
\label{MARS}
\begin{aligned}
    f(\bm x) &= \gamma_0 + \sum_{m=1}^M \gamma_mB_m(\bm x) + \bm\epsilon \\ 
    B_m(\bm x) &= \prod_{i=1}^p\left[s_{im}(x_i - t_{im})\right]_+^{u_{im}},
\end{aligned}
\end{equation}
where $\bm \epsilon$ is a mean-zero error vector, $s_{im} \in \{-1, 1\}$ is called a {\it sign}, $t_{im} \in [0, 1]$ is called a {\it knot} and $u_{im}$ is an indicator equal to $1$ if input $i$ is activated in the $m^{th}$ basis function and equal to $0$ otherwise. The function $[x]_+ = \max(x, 0)$ is the rectified linear unit (ReLU).

A MARS algorithm refers to the process of finding $M$ basis functions ($M$ can be fixed or learned) consisting of signs, knots and indicators so that the linear basis representation is a good approximation of $f(\bm x)$. The original MARS algorithm of \cite{friedman1991}, built basis functions iteratively using a greedy optimization procedure with stochastic search and a set of heuristic rules. In this paper, we use the Bayesian MARS algorithm proposed by \citep{denison1998bayesian} and the improvements discussed by \citep{francom2020bass}. The details of the MARS algorithm are mostly superfluous to the present discussion, so we defer to \citep{francom2020bass} and instead briefly discuss a few relevant points here: 

\begin{itemize}
    \item Training and prediction with MARS involves inverting an $M\times M$ matrix, and therefore the complexity is $\mathcal O(M^3)$. $M$ denotes the number of basis functions, and in practice $M$ scales sub-linearly as a function of $n$. This is a drastic improvement compared to a GP, which has cubic scaling in the size of the training data.
    \item A strength of MARS regression is it's ability to handle a large number of inputs. In most practical applications with a large number of inputs, some inputs are likely to be inert. When dictated by the data, inert variables can be entirely left out of the fitted MARS model. 
    \item The BMARS approach of \citep{francom2020bass} has a few useful properties that we can exploit. First, we usually include the constraint
    $$\sum_{i=1}^p u_{im} \leq J,$$
    which constrains basis functions to include at most $J$ variables. Hence $J$ is the {\it interaction order} and is usually fixed at a small integer (e.g., two or three). This drastically limits the number of integrals that must be solved and bounds the complexity of the entire estimation procedure to $\mathcal O(JM^2 + M^3)$.  Note, again, that $M$ typically scales favorably with both $n$ and $p$, and can be heavily regularized when necessary or desired \citep{francom2020bass}. Additionally, the posterior sampler described in \citep{francom2020bass} makes very small, structured changes in the posterior at each iteration and this can be leveraged for fast updating of the $\bm C$ estimate across MCMC iterations, yielding efficient propagation of model-fit uncertainty to obtain a full posterior for the active subspace. See Sections 3.3 and 6-8 of the SM for additional details. 
\end{itemize}

\section{Methods}
\label{sec:methods}
This section derives a closed form estimator for the matrix $C$ for a flexible class of functions and applies the general result to a MARS regression fit, leading to a practical avenue for active subspace estimation. Specific results are derived for a pragmatic collection of parametric input distributions.  

\subsection{Closed form C in General}
\label{sec:general}
To begin, consider functions of the form  
\begin{equation}
\label{eq:f}
f(\bm x) = \gamma_0 + \sum_{m=1}^M\gamma_m \prod_{i=1}^p h_{im}(x_i), \ \bm x \in \mathcal X \subset \mathbb R^p.
\end{equation}
If the $h_{im}$ functions support univariate approximation then \cref{eq:f} is a universal approximator for any continuous multivariate function over a compact input space $\mathcal X$.  More specifically, if linear combinations of a collection $h_m(x), m=1,2,\ldots$ are dense in $\mathbb C[0,1]$ then linear combinations of $\prod_{i=1}^ph_{im}(x_i)$ are dense in $\mathbb C[0,1]^p$. See \cite{xu2009adaptive, lin1992canonical, shekhtman1982piecewise} for details. Notable special cases of \Cref{eq:f} include MARS/BMARS, radial basis regression \citep{park1993approximation}, additive regression trees \citep{friedman2003multiple} and Nadarya-Watson kernel regression models \citep{bierens1987kernel}. Gaussian process regression, with a separable kernel, can also be viewed as a special case of \cref{eq:f} where $M = n$ \citep{higdon2002space, gramacy2020surrogates,chen2005analytical}.

For now, assume that the distribution on inputs is independent over elements of $\bm x$ so that it can be factored as
\begin{equation}
    \rho(\bm x) = \prod_{i=1}^p \rho_i(x_i),
\end{equation}
although \cref{sec:measure} treats more general cases including finite mixtures of multivariate normals. Noting that the $ij^{th}$ element of $\bm C$ is given as 
$\bm C_{ij} = \int \frac{\partial f(\bm x)}{\partial x_i}\frac{\partial f(\bm x)}{\partial x_j} \rho(\bm x) d\bm x $,
we write 
\begin{equation}
\label{eq:derivatives}
    \begin{aligned}
    \frac{\partial f(\bm x)}{\partial x_i}\frac{\partial f(\bm x)}{\partial x_j} &= \left(\sum_{m_1=1}^M\gamma_{m_1}\frac{dh_{im_1}(x_i)}{dx_i}\prod_{k\neq i}h_{km_1}(x_k)\right)\left(\sum_{m_2=1}^M\gamma_{m_2}\frac{dh_{jm_2}(x_j)}{dx_j}\prod_{\ell\neq j}h_{\ell m_2}(x_\ell)\right) \\[1.5ex]
    &= \begin{dcases}
    \sum_{m_1=1}^M\sum_{m_2=1}^M\gamma_{m_1}\gamma_{m_2}\left(\frac{dh_{im_1}(x_i)}{dx_i}\frac{dh_{im_2}(x_i)}{dx_i}\right)\prod_{k\neq i}h_{km_1}(x_k)h_{k m_2}(x_k), & i = j \\[1.5ex]
    \sum_{m_1=1}^M\sum_{m_2=1}^M\gamma_{m_1}\gamma_{m_2}\left(\frac{dh_{im_1}(x_i)}{dx_i}h_{im_2}(x_i)\right)\left(\frac{dh_{jm_2}(x_j)}{dx_j}h_{jm_1}(x_j)\right) \times \\
    \quad  \prod_{k\not\in \{i, j\}}h_{km_1}(x_k)h_{k m_2}(x_k), & i \neq j.
    \end{dcases}
\end{aligned}
\end{equation}

The assumption $\rho(\bm x) = \prod_{i=1}^p\rho_i(x_i)$ allows for swapping the order of integration and summation in $C_{ij}$ and the resulting integrand can be separated into univariate factors. Thus, the expected value of \cref{eq:derivatives} can be written as a function of univariate integrals as

\begin{equation}
\label{eq:Cij}
    \bm C_{ij} = 
    \begin{dcases}
    \sum_{m_1=1}^M\sum_{m_2=1}^M\gamma_{m_1}\gamma_{m_2} I_3^{(i)}[m_1, m_2]\prod_{k\neq i}I_2^{(k)}[m_1, m_2], & i = j \\[1.5ex]
    \sum_{m_1=1}^M\sum_{m_2=1}^M\gamma_{m_1}\gamma_{m_2}I_1^{(i)}[m_1, m_2]I^{(j)}_1[m_2, m_1] \prod_{k\not\in\{i,j\}}I_2^{(k)}[m_1, m_2], & i \neq j 
    \end{dcases}
\end{equation}
where 
\begin{align}
\label{eq:I1}
    I_1^{(i)}[m_1, m_2] &= \int_{-\infty}^\infty \frac{dh_{im_1}(x)}{dx}h_{im_2}(x) \rho_i(x) dx \\[1.5ex]
\label{eq:I2}
    I_2^{(i)}[m_1, m_2] &= \int_{-\infty}^\infty h_{im_1}(x)h_{im_2}(x) \rho_i(x) dx \\[1.5ex] 
\label{eq:I3}
    I_3^{(i)}[m_1, m_2] &= \int_{-\infty}^\infty \frac{dh_{im_1}(x)}{dx}\frac{dh_{im_2}(x)}{dx} \rho_i(x) dx.
\end{align}
Assembling the entire matrix $\bm C$ requires evaluation of $pM(2M - 1)$ univariate integrals. With suitable choice of $h$ and $\rho$, these integrals can be extremely fast to evaluate. For some choices of $h$ (such as the choice which leads to MARS regression), a large number of these integrals will be zero in practice and, with careful implementation, the actual number of evaluations will be much smaller.

\subsection{Closed Form C with MARS}
\label{sec:resultsmars}
By making the choice
\begin{equation}
\label{eq:gmars}
    h_{im}(x_i) = [s_{im}(x_i - t_{im})]_+^{u_{im}},
\end{equation}
the general results of \cref{eq:Cij,eq:I1,eq:I2,eq:I3} can be applied to the models constructed by the MARS and Bayesian MARS algorithms, and lead to efficient estimation of the active subspace. First, rewrite \cref{eq:gmars} in terms of the indicator function and obtain the derivative
\begin{equation}
\label{eq:gdg}
\begin{aligned}
\chi_{im}(x_i) &= \mathbbm{1}\left(s_{im}(x_i - t_{im}) > 0\right) \\[1.5ex]
h_{im}(x_i) &= [\chi_{im}(x_i)s_{im}(x_i - t_{im})]^{u_{im}} \\[1.5ex]
\frac{dh_{im}(x_i)}{dx_i} &= u_{im}s_{im}\chi_{im}(x_i).
\end{aligned}
\end{equation}
This choice of $h()$ is convenient for \cref{eq:I1,eq:I2,eq:I3}, because $h()$ is piecewise linear (with a single linear component and a constant-zero component) and $dh(x)/dx$ is a step function. This means that the necessary integrands are piecewise polynomial with degree at most $2$. Each integral also contains the product of two indicator functions, which simply have the effect of changing the bounds of integration. To be precise, the integral $\int_{-\infty}^\infty \chi_{im_1}(x_i)\chi_{im_2}(x_i)H(x_i) dx_i$ is equal to $\int_a^b H(x_i)dx_i$ (for arbitrary $H$). In our case, the bounds of integration are given by 
\begin{equation}
\begin{aligned}
    \label{eq:bounds}
    a^{(i)}[m_1, m_2] &= \begin{dcases}
    \max(t_{im_1}, t_{im_2}), & s_{im_1} = +1, \ s_{im_2} = +1 \\
    t_{im_1}, & s_{im_1} = +1, \ s_{im_2} = -1 \\
    t_{im_2},  & s_{im_1} = -1, \ s_{im_2} = +1 \\
    -\infty, & s_{im_1} = -1, \ s_{im_2} = -1
    \end{dcases} 
    \\[1.5ex]
    b_\star^{(i)}[m_1, m_2] &= \begin{dcases}
    \infty, & s_{im_1} = +1, \ s_{im_2} = +1 \\
    t_{im_2}, & s_{im_1} = +1, \ s_{im_2} = -1 \\
    t_{im_1}, & s_{im_1} = -1, \ s_{im_2} = +1 \\
    \min(t_{im_1}, t_{im_2}), & s_{im_1} = -1, \ s_{im_2} = -1
    \end{dcases} \\[1.5ex]
    b^{(i)}[m_1, m_2] &= \max\left\{b_\star^{(i)}[m_1, m_2], a^{(i)}[m_1,m_2]\right\}.
\end{aligned}
\end{equation}
Because the relevant integrands are piecewise polynomials with degree at most two, we only need to be able to evaluate integrals of the form 
\begin{equation}
\label{eq:truncmoments}
    \xi(r | a, b, \rho_i) = \int_a^b x^r \rho_i(x) dx, \ r=0, 1, 2.
\end{equation}
That is, $\xi(r | a, b, \rho_i)$ is the $r^{th}$ truncated moment with respect to the measure $\rho_i$ and the truncation interval $(a, b)$. This equation will be a primary motivator during our discussion of reasonable choices of prior in the next subsection. Given \cref{eq:bounds,eq:truncmoments}, we can rewrite the equations for $I_\ell^{(i)}[m_1, m_2]$ for this important special case. For notational brevity, we write $\xi(r)$ as a stand-in for $\xi(r | a^{(i)}[m_1, m_2], b^{(i)}[m_1, m_2], \rho_i)$. 

\begin{align}
\label{eq:I1mars}
I_1^{(i)}[m_1, m_2] &= s_{im_1}s_{im_2}
\begin{dcases}
u_{im_1}\left(\xi(1) - t_{im_2}\xi(0)\right), & u_{im_2} = 1 \\[1.5ex]
u_{im_1}\xi(0) , & u_{im_2} = 0
\end{dcases} \\[2.5ex]
\label{eq:I2mars}
I_2^{(i)}[m_1, m_2] &= s_{im_1}s_{im_2}
\begin{dcases}
\xi(2) - (t_{im_1} + t_{im_2})\xi(1) + t_{im_1}t_{im_j}\xi(0), & u_{im_1} = 1, \ u_{im_2} = 1 \\[1.5ex]
\xi(1) - t_{im_1}\xi(0), & u_{im_1} = 1, \ u_{im_2} = 0 \\[1.5ex]
\xi(1) - t_{im_2}\xi(0), & u_{im_1} = 0, \ u_{im_2} = 1 \\[1.5ex]
\xi(0), & u_{im_1} = 0, \ u_{im_2} = 0 
\end{dcases} \\[2.5ex]
\label{eq:I3mars}
I_3^{(i)}[m_1, m_2] &= u_{im_1}u_{im_2}s_{im_1}s_{im_2}\xi(0) 
\end{align}

In summary, estimation of the active subspace of a function $f$ with respect to the independence prior $\bm x \sim \rho(\bm x) = \prod_{i=1}^p\rho_i(x_i)$ proceeds as follows. Given pairs $(\bm x_i, y_i)$ such that $y_i = f(\bm x_i) + \epsilon_i$, follow \cite{francom2020bass} to construct a surrogate of $f$ which can be written in the form of \cref{eq:f,eq:gmars}. Using the fitted MARS model, construct the $M \times M$ (where $M$ is the number of basis functions of the MARS model) matrices $a^{(i)}, b^{(i)}, I_1^{(i)}, I_2^{(i)}, I_3^{(i)}$ for $i=1,\ldots p$, requiring $\mathcal O(pM^2)$ memory. These matrices are easily  constructed, so long as the truncated moments (\cref{eq:truncmoments}) are readily evaluated. Then construct $\bm C$ using \cref{eq:Cij}. A substantial amount of time is saved by recognizing that $a^{(i)}, b^{(i)}, I_2^{(i)}, I_3^{(i)}$ and $\bm C$ are symmetric. Finally, if there are memory concerns regarding the storage of these matrices, \cref{eq:Cij} can be used to construct $\bm C$ {\it in situ}, at the cost of re-computing each integral $p(p+1)/2$ times. Pseudocode for this procedure is given in Section 1 of the SM. 

\subsection{The Choice of Measure}
\label{sec:measure}
In the active subspace literature, uniform probability measure is the most common choice for $\rho$. This is convenient because it only requires the subject matter expert to specify a lower and upper bound for the reasonable values of each input $x_i$. If no such sensible bounds exist, Gaussian measures represent a natural starting point \citep{wycoff2021sequential}. In the presence of stronger information, more suitable results can be obtained by using a measure more tailored to the problem of interest. For example, flexible parametric distributions such as the Beta (for bounded inputs), Gamma (for semi-bounded inputs) and mixtures of Gaussians (for unbounded inputs) may be more appropriate, in certain settings, compared to uniform measure. In other problems, there may exist important covariance structure between inputs or possibly even strict constraints (e.g., $x_1 < x_2$). This section explores use of input distributions beyond independent uniforms, demonstrating compatibility with many desirable choices of prior. 

First, we state a useful Theorem that we rely upon in the remainder of this section.
\begin{theorem}
\label{thm1}
Let $\bm z = \bm A\bm x + \bm b$ and take $\tilde f(\bm z) = f(\bm x)$. If $\bm A$ is invertible, then we have
$$\bm C = \bm A^\intercal \tilde{\bm C} \bm A$$
where $\bm C$ is the $C$ matrix taken with respect to $f$ and $\rho_x$ and $\tilde{ \bm C}$ is taken with respect to $\tilde f$ and $\rho_z(z) = |\bm A|^{-1} \rho_x(\bm A^{-1}(\bm z - \bm b))$.
\end{theorem}
Thus, the $C$ matrix with respect to linearly transformed inputs (i.e.~$\tilde{ \bm C}$) is easily converted to the $C$ matrix of the native inputs. A proof of \cref{thm1} is given in Section 2 of the SM.

\subsubsection{Independent Inputs}
\label{sec:univariate}
This is the simplest case: $\rho(\bm x) = \prod_{i=1}^n\rho_i(x_i)$ with the $\rho_i(x_i)$ being simple parametric distributions. For instance, we commonly take $x_i \stackrel{\text{iid}}{\sim} \text{Unif}(0, 1)$, and \cref{eq:truncmoments} becomes
\begin{equation}
\label{eq:unif}
\begin{aligned}
    \tilde a &= \max\{a, 0\}, \quad\quad \tilde b = \min\{b, 1\} \\
    \xi_\text{Unif}(r | a, b, \rho_i) &= \int_{\tilde a}^{\tilde b} x^r dx = \frac{\tilde b^{r+1} - \tilde a^{r+1}}{r + 1}.
\end{aligned}
\end{equation}
The case of an arbitrary uniform distribution follows from \cref{thm1} with $\bm A$ and $\bm b$ taken to shift and scale the inputs to $[0,1]$ with recovery of $C$ according to the Theorem. The uniform case can be generalized by specifying a flexible two-parameter Beta distribution for each input, $x_i \stackrel{\text{ind}}{\sim} \text{Beta}(\alpha_i, \beta_i)$.
\begin{equation}
\label{eq:beta}
\begin{aligned}
    \tilde a &= \max\{a, 0\}, \quad\quad \tilde b = \min\{b, 1\} \\
    \xi_\text{Beta}(r | a, b, \rho_i) &=  \frac{B_{\tilde b}(\alpha + j, \beta) - B_{\tilde a}(\alpha+j, \beta)}{B_1(\alpha, \beta)},
\end{aligned}
\end{equation}
where $B_x(\alpha, \beta) = \int_0^xt^{\alpha-1}(1-t)^{\beta-1}dt$ is the {\it incomplete Beta function}.

It is not uncommon for a computer model to have inputs which must be positive. In these situations, it can be useful to specify a measure for these inputs with strictly positive support. For instance, if $\rho_i(x_i)$ is taken to be a Gamma distribution with shape $\alpha$ and rate $\beta$, then we obtain
\begin{equation}
\label{eq:gamma}
    \xi_\text{Gamma}(r | a, b, \rho_i) = \frac{\Gamma_{\beta b}(\alpha+j) - \Gamma_{\beta a}(\alpha+j)}{\beta^j\Gamma_\infty(\alpha)},
\end{equation}
where $\Gamma_x(\alpha) = \int_0^xt^{\alpha-1}e^{-t}dt$ is the {\it lower incomplete Gamma function}.

Next, we consider the useful and familiar Gaussian distribution with mean $\mu$ and standard deviation $\sigma$. For additional flexibility, we also consider truncation to the interval $(\tau_0, \tau_1)$. Since $\xi$ already involves truncated moments, using a truncated version of any distribution involves little extra effort. Evaluating moments of the truncated normal can be a challenging problem and many standard formulations lead to {\it catastrophic cancellation}, a classic source of floating point error  \citep{truncNorm}. The form presented below is carefully constructed to be robust in the ways which are important here. Namely, we have found the following to produce estimates of $\bm C$ with numerically non-negative eigenvalues, even in challenging high-dimensional problems. 
\begin{equation}
\label{eq:xitnorm}
\begin{aligned}
    \tilde a &= \max\{a, \tau_0\} \\
    \tilde b &= \max\{\tilde a, \min\{b, \tau_1\}\} \\
    \Delta_1 &= \Phi\left(\frac{ \tilde b - \mu}{\sigma}\right) - \Phi\left(\frac{\tilde a - \mu}{\sigma}\right),
    \qquad 
    \Delta_2 = \Phi\left(\frac{ \tau_1 - \mu}{\sigma}\right) - \Phi\left(\frac{\tau_0 - \mu}{ \sigma}\right) \\
    Z_1 &= \phi\left(\frac{\tilde a - \mu}{\sigma}\right) - \phi\left(\frac{\tilde b - \mu}{\sigma}\right), \qquad
    Z_2 = \Delta_1 -  \frac{\tilde a - \mu}{\sigma}\phi\left(\frac{\tilde a - \mu}{\sigma}\right) - \frac{\tilde b - \mu}{\sigma}\phi\left(\frac{\tilde b - \mu}{\sigma}\right) \\[2ex]
    &\xi_\text{TNorm}(r | a, b, \rho_i) = \frac{\Delta_1 \mu^r}{\Delta_2} 
    + \frac{1}{\Delta_2} \times \begin{dcases}
    0, & r= 0 \\
    \sigma Z_1, & r= 1 \\
    2\sigma\mu Z_1 + \sigma^2 Z_2, & r=2,
    \end{dcases} 
\end{aligned}
\end{equation}
where $\phi$ and $\Phi$ denote the density and distribution functions of the standard normal, respectively.  

\Cref{eq:unif,eq:beta,eq:gamma,eq:xitnorm} are not meant to represent an exhaustive list of measures which are compatible with \cref{eq:truncmoments}, but rather to demonstrate that this approach is extremely flexible when it comes to specifying a measure which is independent across the inputs. Other choices, such as Student's-$t$ and log-Normal distributions (with or without truncation) may be useful for other applications. 

\subsubsection{Multivariate Gaussian Inputs}
\label{sec:mvnorm}
In complex physical models, it is common to assume or estimate (e.g., through statistical calibration) a covariance structure across inputs. In this section, we build on the previous discussion to construct an estimator for $\bm C$ when $\rho(\bm x)$ is a multivariate normal distribution. 

If $\bm x \sim N(\bm \mu, \bm\Sigma)$, then $\bm z = \bm\Sigma^{-1/2}\left(\bm x - \bm \mu\right) \sim N(\bm 0, \bm I)$ where $\bm I$ is the identity matrix and $\bm\Sigma^{-1/2}$ is a square root of $\bm\Sigma^{-1}$. The measure for the transformed variables $\bm z$ can now be factored $\rho_z(\bm z) = \prod_{i=1}^p\rho_z(z_i)$, and we can obtain $\bm C_z$ using the methods described in \cref{sec:univariate}. Taking $\bm A = \bm \Sigma^{-1/2}$ and $\bm b = -\bm \Sigma^{-1/2} \bm \mu$, we can leverage \cref{thm1} to obtain the $C$ matrix for the original inputs $\bm x$ as 
\begin{equation}
\label{eq:Cmvn}
    \bm C = \left(\bm \Sigma^{-1/2}\right)^\intercal \bm C_z \bm \Sigma^{-1/2}
\end{equation}

We note that this implies that the emulator must be trained to regress y on $\bm z$ rather than on $\bm x$. While this will lead to a different fitted form, the linear transformation of the inputs is unlikely to make the model more difficult to fit, since emulators are flexible nonlinear models. 

\subsubsection{Mixture Distributions}
\label{sec:indmix}
\label{sec:genmix}
For additional flexibility and generality, we now consider the use of mixture distributions which can be used to approximate practically any measure $\rho$ which may be of interest. There are two primary ways in which mixture distributions can be leveraged in our framework. In the first case, one may be willing to assume independence between the input variables $x_i$ and $x_{i'}$, but we desire a more flexible framework for specifying the marginal distributions of each input. This additional layer of flexibility allows for multi-modal or arbitrarily skewed distributions to be approximated. In this first case, we assume again that $\rho(\bm x) = \prod_{i=1}^p\rho_i(x_i)$, but now allow $\rho_i(x_i)$ to be a finite mixture distribution
\begin{equation}
    \rho_i(x_i) = \sum_{\ell=1}^{L_i} \omega_{i\ell}\rho_{i\ell}(x_i), \ \sum_{\ell=1}^L\omega_{i\ell} = 1
\end{equation}
and \cref{eq:truncmoments} becomes
\begin{equation}
\label{eq:mixindependent}
    \xi^\prime(r | a, b, \rho_i) = \int_a^b x^r \sum_{\ell=1}^{L_i} \omega_{i\ell}\rho_{i\ell}(x_i)dx = \sum_{\ell=1}^{L_i}\omega_{i\ell}\xi(r | a, b, \rho_{i\ell}).
\end{equation}
In other words, the finite mixture distribution can be handled trivially, so long as each component of the mixture can be handled. Of course, this can have a negative impact on run-time if $L_i$, the number of mixture components, is very large. The asymptotic cost increases by a factor of $\frac{1}{p}\sum L_i$, since $L_i$ integrals are now required for each input and basis function combination. 

A second, more general, way of thinking about mixture distributions is to consider input distributions of the form
\begin{equation}
\label{eq:mixgen}
    \rho(\bm x) = \sum_{\ell=1}^L\omega_\ell \rho_\ell(\bm x).
\end{equation}
Here, it is easy to show that
\begin{equation}
\label{eq:Cmixgen}
    \bm C = \sum_{\ell=1}^L w_\ell \bm C_\ell
\end{equation}
where $\bm C_\ell$ is the $C$ matrix with respect to $\rho_\ell(\bm x)$. In order to compute $\bm C_\ell$, we require that each $\rho_\ell(\bm x)$ is either (i) a multivariate normal distribution or (ii) a product of $p$ independent univariate distributions (e.g., independent components). In cases where either approach can be used, \cref{eq:mixindependent} requires less overhead and no extra memory but \cref{eq:Cmixgen} can be parallelized more easily. 

Immediately, two important use-cases come to mind. First, consider the case where $\rho(\bm x)$ is not analytically available, and is only provided through a set of observations $\bm x^\prime_1, \bm x^\prime_2, \ldots \bm x^\prime_N$. This scenario could arise, for instance, when $\rho(\bm x)$ represents the output of a model calibration procedure \citep{kennedy2001bayesian}. The distribution $\rho(\bm x)$ can be approximated using a mixture of multivariate normal distributions \citep{redner1984mixture}, and written in the form of \cref{eq:mixgen}. Secondly, mixtures of this form are useful for approximating distributions with strict mathematical constraints on the support, which we are otherwise unable to accommodate in this framework. Examples are given for both of these scenarios in \cref{sec:examples}. Note that computation of each $\bm C_\ell$ is trivially parallelizable, so that computing $\bm C$ using \cref{eq:Cmixgen} need not be much costlier than computing a single $\bm C_\ell$.

\section{Examples}
\label{sec:examples}
This section demonstrates the efficiency and accuracy of active subspace estimation with Bayesian MARS for a variety of problems. In particular, the MARS-based method is shown to be more efficient and accurate than the GP based approach of \citep{wycoff2021sequential} when the number of inputs $p$ is large. When $p$ is not large, the difference between the estimators is small, but our method scales much better with the size of the training data. 

We also apply these methods to two applications of genuine physical interest, showing the flexibility of the prior specification and the ability to easily handle a large number of inputs. Additional examples can be found in Section 3 of the SM. 

All examples are conducted with the {\it concordance} R package, which can be found at \url{https://github.com/knrumsey/concordance}. Scripts to reproduce the examples found in \cref{sec:polynomial} are available at \url{https://github.com/knrumsey/ASM-BMARS-Examples}.

\subsection{A Simple Polynomial Benchmark}
\label{sec:polynomial}
For illustrative purposes, we begin with a simple case for which the true $\bm C$ matrix can be obtained analytically. Consider the function 
\begin{equation}
    \label{eq:simple}
    f(\bm x) = x_1^2 + x_1 x_2 + \frac{1}{9}x_2^3
\end{equation}
First, we consider the simple case where $\bm x$ is uniform over the unit hypercube; i.e., $\rho_1(\bm x) = \mathbbm{1}(0 < x_1 < 1)\mathbbm{1}(0 < x_2 < 1)$. Applying \cref{eq:C} directly gives the $\bm C$ matrix
\begin{equation}
\label{eq:Cex1}
    \bm C_{\rho_1} = \frac{1}{45}
  \begin{bmatrix}
    120 & 50  \\[1.5ex]
    50 & 21 
  \end{bmatrix} .
\end{equation}
Although $f(\bm x)$ is influenced by just two inputs, it can be informative to add a collection of {\it inert} inputs, taking $\bm x = (x_1, x_2, \ldots x_p)^\intercal$. In this case, the $\bm C$ matrix is $p \times p$ having \cref{eq:Cex1} as an upper left block and zeros elsewhere. 

To estimate $\bm C$, we generate training examples $(\bm x_1, f(\bm x_1)), (\bm x_2, f(\bm x_2)), \ldots (\bm x_n, f(\bm x_n))$ using a Latin hypercube design over $[0, 1]^p$ \citep{park1994optimalLHS}, and fit a surrogate model to the resulting data. For a MARS surrogate, $\bm C$ is estimated using the results of \cref{sec:methods} and for a Gaussian process surrogate we use the method of \citep{wycoff2021sequential}. Effectiveness of each approach is measured by (i) cost (in seconds) of fitting the surrogate model, (ii) cost (in seconds) of estimating $\bm C$, (iii) error in the estimate of $\bm C$, and (iv) error in the estimate of the first active direction. Error in the estimate of $\bm C$ is quantified using the Euclidean norm (also the Frobenius norm), defined as
\begin{equation}
    \sqrt{\sum_{i=1}^p\sum_{i=1}^p \frac{\left(\bm C_{ij} - \hat{\bm C}_{ij}\right)^2}{p^2}} = \frac{1}{p}\sqrt{\text{Tr}\left[\left(\bm C - \hat{\bm C}\right)\left(\bm C - \hat{\bm C}\right)^\intercal\right]},
\end{equation}
where $\bm C$ is the true matrix and $\hat{\bm C}$ denotes the estimate. The first active direction for $f$ with respect to $\rho_1$ is $\bm w_1 \approx [0.923 \ \ 0.385 \ \ 0 \ \ \ldots \ \ 0]^\intercal \in \mathbb R^p$ and estimation error is measured by the standard Euclidean vector norm. 

\begin{figure}[t]
\centering
\includegraphics[width=\textwidth]{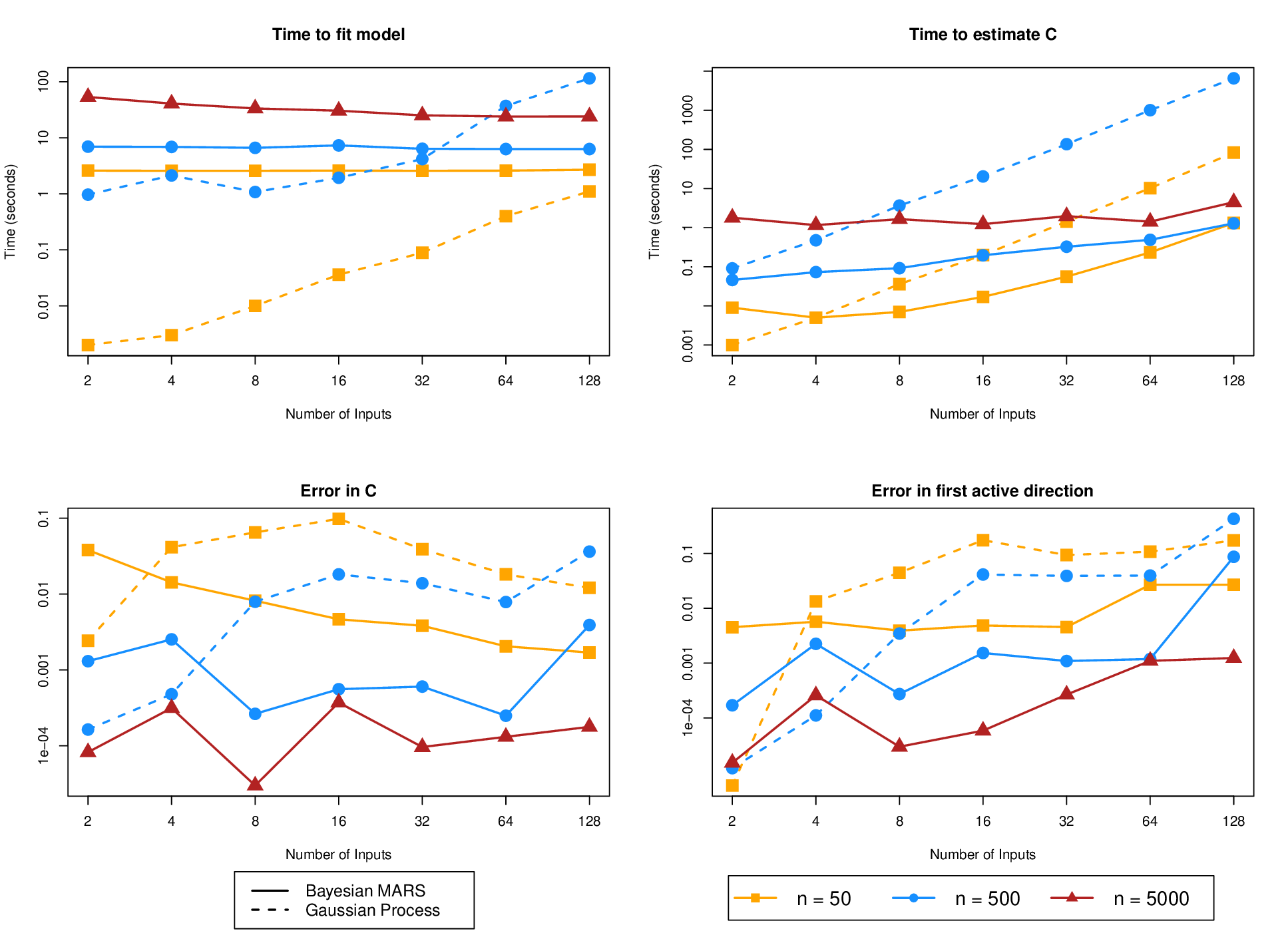}
\caption{Results of the simulation study for \cref{eq:simple} with a standard uniform prior. The standard Gaussian process cannot be easily fit for $n = 5000$ so results are omitted.}
\label{fig:example1}
\end{figure}

To conduct the analysis, we generate Latin hypercube designs of size $n = 50$, $500$ and $5000$ with dimensions $p = 2^d (d=1,\ldots 7)$ ({\it Note:} to reduce variance in the results, a single design is generated for each $n$ using $p_\text{max} = 2^7$ and surrogates are fit for each $p$ using just the first $p$ inputs of each design). The Bayesian MARS surrogate was fit in the open-source R programming language using the \texttt{BASS::bass()} function \citep{francom2020bass} and the GP surrogate was fit using the \texttt{hetGP::mleHomGP()} function \citep{binois2021hetgp}. For the respective cases, $\bm C$ was estimated using the \texttt{concordance::C\_bass()} and \texttt{activegp::C\_GP()} functions \citep{activegp}.  

The results of this simulation study are fully presented in \cref{fig:example1}. As expected, the computational cost of the GP-surrogate approach drastically increases as either $p$ or $n$ increase. For example, in the medium-difficulty case with $n=500$ and $p=64$, the GP based approach required $37$ seconds to fit the surrogate and nearly $17$ minutes to estimate $\bm C$. For the case where $n=5000$ and $p=128$, we approximate that estimating $\bm C$ with the GP-based approach would have taken weeks (for this reason, we did not test the $n=5000$ case for GPs at all). The MARS-based approach had much better scaling. For the high-difficulty case with $n=5000$ and $p=128$, we required just $24$ seconds to fit the surrogate and $4.5$ seconds to estimate $\bm C$. 

For very low-dimensional input spaces ($p=2$), the GP-based approach led to more precise estimates of both $\bm C$ and $\bm w_1$ compared to the MARS approach for equivalent $n$. On the other hand, estimates based on MARS are more accurate for moderate and large $p$ ($p \geq 4$ in some cases, and $p \geq 8$ in all cases). Moreover, the estimators produced by MARS can be improved by adding training data (e.g., see $n=5000$ case in \cref{fig:example1}) while the GP cannot easily take advantage of these additional runs for computational reasons.

\begin{figure}[t]
\centering
\includegraphics[width=\textwidth]{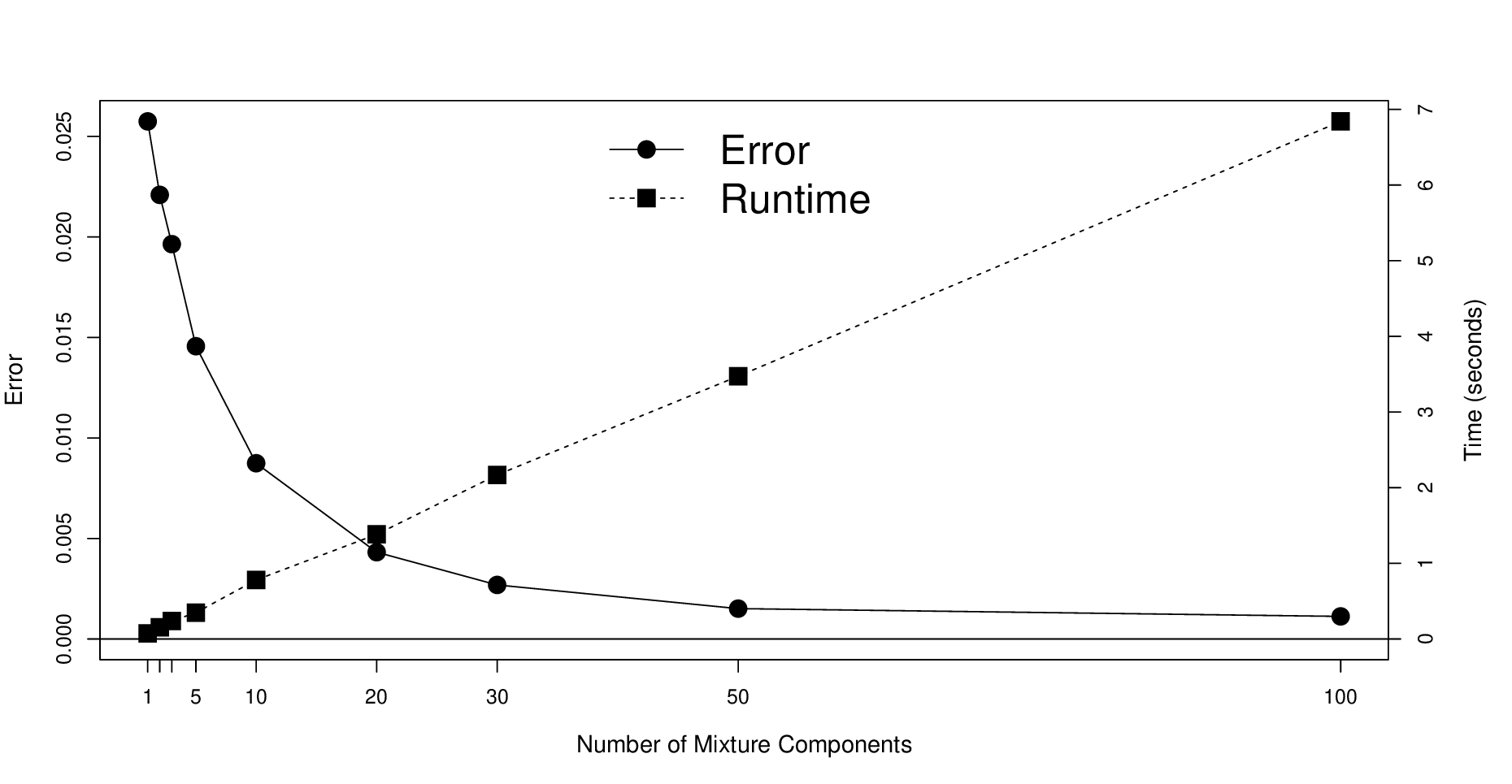}
\caption{The left vertical axis shows the error of the approximation of $\bm C$ (solid line, circles) as a function of $L$. The right vertical axis shows the time taken (in seconds) to obtain the approximation.}
\label{fig:example1b}
\end{figure}

\subsubsection{A Linear Constraint}
Let us again consider the function of \cref{eq:simple} (with $p = 2$), but this time we will assume the existence of the constraint $x_1 > x_2$. In particular, suppose we want to find the active subspace of $f$ with respect to the prior
\begin{equation}
    \rho_2(x_1, x_2) = 2\cdot \mathbbm 1(0 < x_2 < x_1 < 1),
\end{equation}
which represents a uniform distribution over the lower right half of $[0, 1]^2$. As before, we can determine the $\bm C$ matrix analytically using \cref{eq:Cij}
\begin{equation}
    \bm C_{\rho_2} = \frac{1}{540} 
    \begin{bmatrix}
    1710 & 741 \\[1.5ex]
    741 & 322
\end{bmatrix} .
\end{equation}
This measure cannot be handled directly in our framework because the distribution does not factor into independent components, and no linear transformation of $\bm x$ will help here. Instead, we approximate $\rho_2(\bm x)$ with a mixture of $L$ uniform distributions as
\begin{equation}
\begin{aligned}
    \rho_2(x_1, x_2) &\approx \sum_{\ell=1}^L\omega_\ell \frac{(L+1)^2}{L+1-\ell}\mathbbm{1}\left(\frac{\ell}{L+1} < x_1 < 1\right)\mathbbm{1}\left(\frac{\ell-1}{L+1} < x_2 < \frac{\ell}{L+1}\right) \\[1.2ex]
    &= \frac{2(L+1)}{L}\sum_{\ell=1}^L\mathbbm{1}\left(\frac{\ell}{L+1} < x_1 < 1\right)\mathbbm{1}\left(\frac{\ell-1}{L+1} < x_2 < \frac{\ell}{L+1}\right)
\end{aligned}
\end{equation}
The mixture weights are proportional to the size of each region and are computed as $\omega_\ell \propto \frac{L+1-\ell}{(L+1)^2}$, subject to $\sum_{\ell=1}^L\omega_\ell = 1$. From here, $\bm C$ is estimated using \cref{eq:unif,eq:mixgen}. Although the accuracy of the approximation improves as $L$ grows large, the computation also grows linearly in $L$ (though parallel computing can easily offset this cost). Finally, note that for certain constructions, this mixture-based procedure becomes increasingly similar to the standard Monte Carlo approach as $L \rightarrow \infty$. The error of the approximation and the time required are shown as a function of $L$ in \cref{fig:example1b}. 

Although we perform a similar analysis in a higher-dimension for a real application in \cref{sec:ptw}, finding a suitable approximation with independent components becomes increasingly difficult as $p$ grows and as the constraints become more complex. See Section 4 of the SM for a practical approach. 

\subsubsection{Capturing Low-Dimensional Structure}

We rely on the simple polynomial function one final time to illustrate the low-dimensional structure which can be captured by active subspaces. In this example, take $\bm x = (x_1, \ldots, x_6)$ and define a new computer model $g: \mathbb R^6 \rightarrow\mathbb R$ as 
\begin{figure}[h]
\centering
\begin{subfigure}{.45 \textwidth}
  \centering
  \includegraphics[width=.95\textwidth]{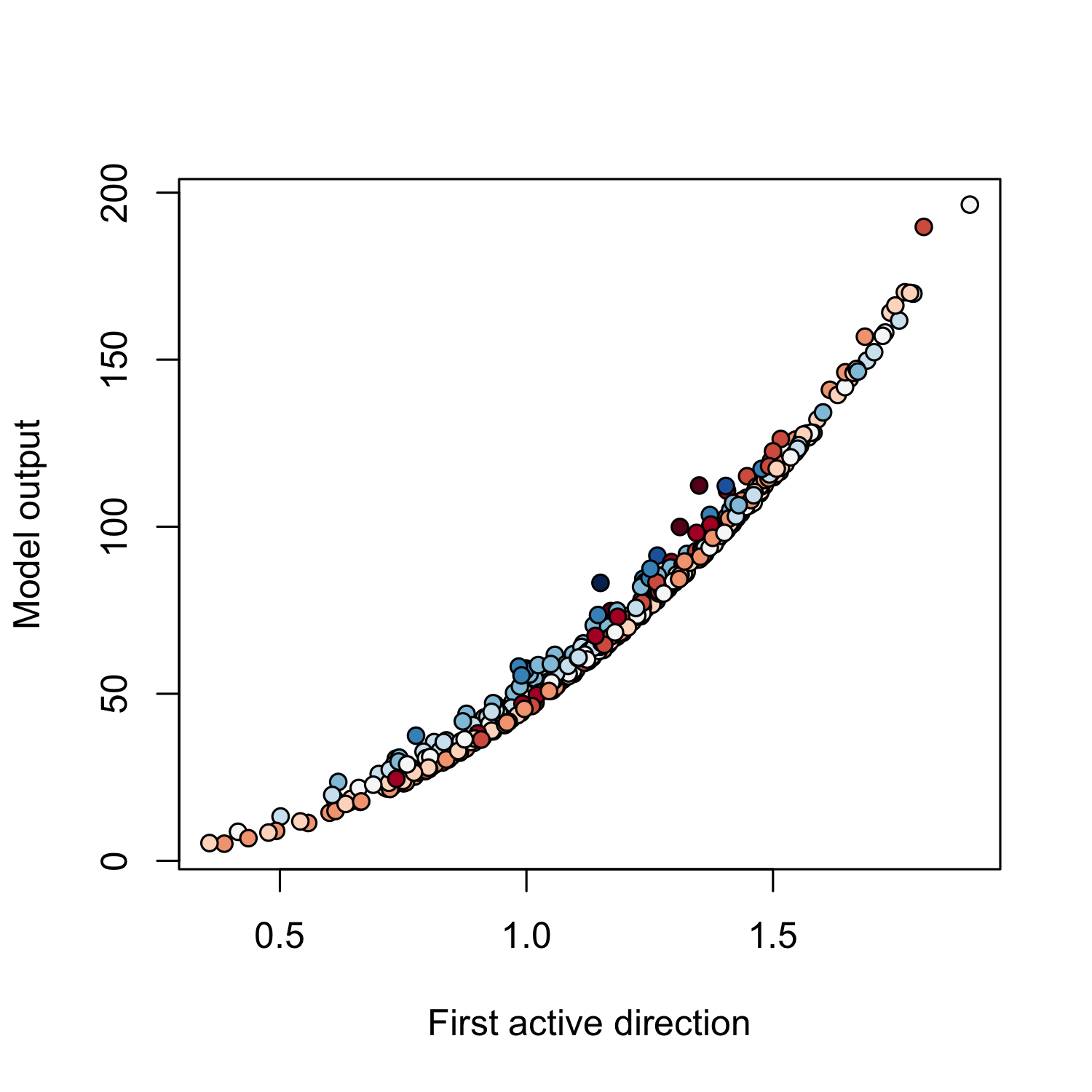}
  \caption[width=0.8\textwidth]{Model output vs. the data projected onto the first active direction. Non-linear structure is present, but not fully captured by just one active direction. }
  \label{fig:poly_2d}
\end{subfigure}%
\quad
\begin{subfigure}{.45\textwidth}
  \centering
  \includegraphics[width=.95\textwidth]{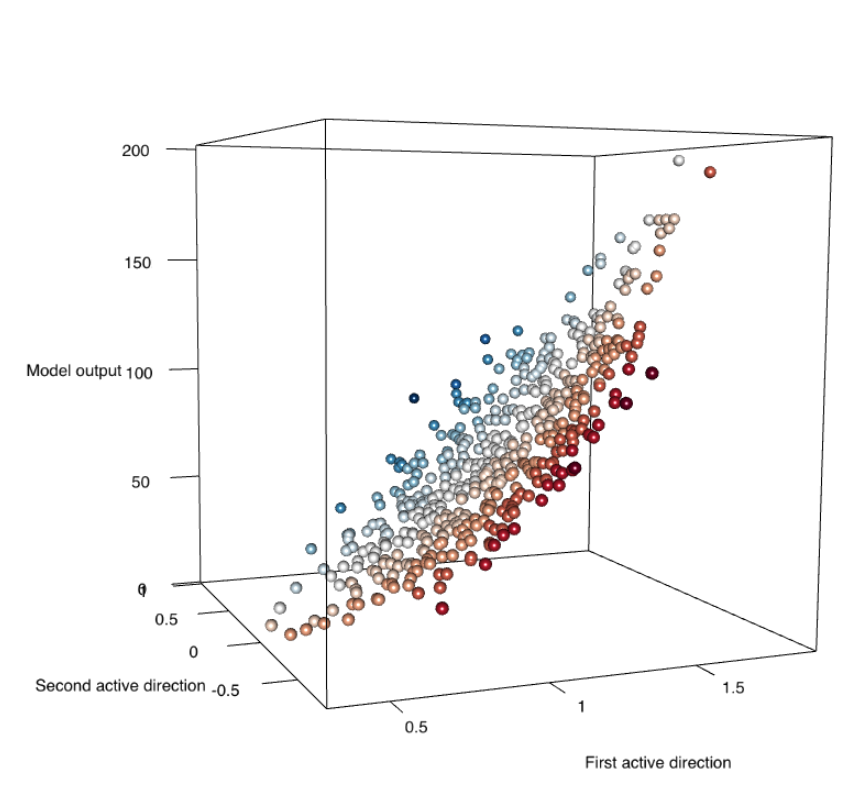}
  \caption{Model output vs. the data projected onto the first and second active directions. The computer model output can be almost completely captured by two active directions. }
  \label{fig:poly_3d}
\end{subfigure}
\caption{Low-dimensional representations of $g$. Points are shaded according to the value of the second active variable for better visualization. }
\label{fig:poly_scatter}
\end{figure}
\begin{table}[!htbp] \centering 
  \caption{The first row gives the square root of each eigenvalue $(i=1,\ldots, 6)$ which is estimated based on our proposed method. The second row reports the percentage of the standard deviation in $g$ which is accounted for by a statistical model fitted using the first $i$ active projections ($i=1,\ldots, 6$).} 
  \label{tab:poly} 
\begin{tabular}{@{\extracolsep{5pt}} lcccccc} 
\\[-1.8ex]\hline 
\hline \\[-1.8ex] 
& \multicolumn{6}{c}{Active Directions} \\[1.2ex] 
& 1 & 2 & 3 & 4 & 5 & 6 \\ \hline \\[-1.8ex]
Square Root Eigenvalue & $1$ & $0.125$ & $0.057$ & $0.046$ & $0.037$ & $0.036$ \\ 
Percent of SD Explained & $91.057$ & $98.368$ & $99.025$ & $99.393$ & $99.448$ & $99.452$ \\ 
\hline \\[-1.8ex] 
\end{tabular} 
\end{table} 
$g(\bm x) \equiv f(\bm x \bm t_1, \bm x \bm t_2),$ with $f$ from \cref{eq:simple}, $\bm t_1 = (12, 9, 6, 1, 1, 1)/3$ and $\bm t_2 =(1, 1, 1, 15, 6, 6)/3$.
 Next, (i) generate a space filling design over $[0,1]^6$ using a Latin hypercube with $n=500$ observations, (ii) fit a Bayesian MARS model to the resulting data and (iii) estimate $\bm C$ using our proposed method. Although $g$ is a function in $6$ dimensions and none of the $6$ inputs are inert, the active subspace can be used to construct a lower-dimensional representation of the data. \Cref{fig:poly_2d} shows the non-linear structure that arises when plotting the model output against the data projected onto the first active direction ($\bm X\bm w_1$). Although the one-dimensional structure is apparent, the output cannot be fully explained by just one active direction. \Cref{fig:poly_3d} shows a 3-dimensional scatterplot of $g(\bm x)$ against the first two active directions. Note that nearly all of the unexplained variability in $g$ is now captured by the reduced data and this is supported by the eigenvalues, shown in \cref{tab:poly}, where the second eigenvalue is not close enough to $0$ in magnitude to justify excluding the second active direction. To illustrate further, we fit statistical regression models (Bayesian MARS) for $g$ using the $d$-dimensional projection (for $d=1,\ldots, 6$) and calculate the percentage of the standard deviation which is accounted for by the statistical model (see \cref{tab:poly}). Each additional dimension beyond 2 yields diminishing returns for the predictive accuracy of the statistical model.

\subsection{A Material Strength Model}
\label{sec:ptw}
In this section, we demonstrate the utility of our approach on an important application in materials science, where we seek to discover the active subspace of a material strength model with respect to two non-trivial input distributions.

The Preston-Tonks-Wallace (PTW) model is a materials strength model which describes the plasticity of a given metal at various temperatures and strain rates \citep{preston2003model}. The equations governing the PTW model are too lengthy to show here, but full details can be found in \cite{fugate2005hierarchical} and \cite{price2013validation}. The PTW model is a complex, highly non-linear function of $p=11$ inputs, corresponding to various material properties, universal constants and measured values. The impact of each input can vary for different metals, temperatures, strain rates and values of the other inputs, so specification of an appropriate measure is of paramount importance. 

\begin{table}
    \centering
    \begin{tabular}{@{\extracolsep{5pt}}lllll}
    \\[-1.8ex]\hline 
\hline \\[-1.5ex] 
        Input & Parameter & Lower bound & Upper bound & Description \\
\hline \\[-1.5ex] 
    $x_1$    & $\theta$ & $0.0001$ & $0.2$ & Strain hardening rate  \\
    $x_2$    & $p$ & $0.0001$ & $5$ &  Strain hardening constant \\
    $x_3$    & $s_0$ & $0.0001$ & $0.05$ &  Saturation stress at $0$ K \\
    $x_4$    & $s_\infty$ & $0.0001$  & $0.05$ &  Saturation stress at melting temperature \\
    $x_5$    & $\kappa$ & $0.0001$ & $0.5$ & Constant of temperature dependence  \\
    $x_6$    & $-\log\gamma$ & $9.21$ & $13.82$  & Constant of strain rate dependence ($\gamma$)  \\
    $x_7$    & $y_0$ & $0.0001$  & $0.05$ & Yield stress constant at $0$ K  \\
    $x_8$    & $y_\infty$ & $0.0001$ & $0.01$ & Yield stress constant at melting temperature  \\
    $x_9$    & $y_1$ &$0.001$  & $0.1$ & Medium strain rate constant  \\
    $x_{10}$ & $y_2$ & $0.33$ & $1$ & Medium strain rate exponent  \\
    $x_{11}$ &  & $0.99$ & $1.1$ & Inert input  \\[1.2ex] \hline
    \end{tabular}
    \caption{Inputs to the PTW model for Ti64 and the corresponding marginal parameter ranges for Ti64. See \citep{price2013validation} for details about each parameter.}
    \label{tab:PTW_pars}
\end{table}

For our purposes, we will use the output of a PTW computer experiment conducted for Ti64, a titanium alloy. The response of interest is the yield stress at a strain of $1.0$, a temperature of $480$ K and a strain rate of $1.1\times 10^6$ per second. Marginal input ranges for each of the $11$ parameters are given in \cref{tab:PTW_pars}. In addition to these marginal ranges, the PTW model must also respect the constraint
\begin{equation}
\label{eq:constraint}
    x_8 \leq \min\{x_4, x_7\} \leq \max\{x_4, x_7\} \leq x_3 \leq x_9.
\end{equation}
This inequality defines a region $\mathcal X \subset \mathbb R^{11}$ which represents the support of $\bm x$. The first measure of interest is the uniform distribution over $\mathcal X$. Since we cannot compute $\bm C$ with respect to this measure directly, we propose to approximate $\mathcal X$ with the union of disjoint hyperrectangles (boxes), and then to approximate $\bm C$ with respect to the mixture of uniforms over these boxes. 
\begin{equation}
    \begin{aligned}
    \rho_\text{prior}(\bm x)  &= \frac{\mathbbm{1}\left(\mathcal X\right)}{\text{Volume}\left(\mathcal X\right)} \\
    &\approx \sum_{\ell = 1}^L\omega_\ell \frac{\mathbbm{1}\left(\bm x \in \mathcal R_\ell\right)}{v_\ell}
    \end{aligned}
\end{equation}
Each box can be represented by a vector of lower bounds $a_j$  and a vector of upper bounds $b_j$ ($j=1,2\ldots 11$). The mixture weights are equal to $v_j/\sum_{\ell=1}^lv_\ell$ where $v_j = \prod_{j=1}^{11}(b_j - a_j)$. Each component of the mixture $\rho_\ell(\bm x) = \mathbbm{1}\left(\bm x \in \mathcal R_\ell\right)/v_\ell$ can be handled easily using \cref{sec:univariate} and the combined $\bm C$ can be found using \cref{eq:Cmixgen}. To find a suitable approximation $\mathcal X \approx \bigcup_{\ell=1}^L R_l$, we employ an approach which is inspired by recursive partitioning \citep{therneau1997introduction}. A detailed description can be found in Section 4 of the SM. The procedure stops when it can no longer recursively add a disjoint hyperrectangle inside of $\mathcal X$ with volume greater than $10^{-12}$, which leads to an approximation consisting of $L = 478$ disjoint boxes. 

The second measure of interest is the posterior distribution of the inputs which results from a Bayesian model calibration analysis \citep{kennedy2001bayesian}. The details are not important here, except that this posterior is unavailable analytically. Instead, we are given a sample of $1500$ draws $\bm x^\prime_1, \ldots \bm x^\prime_{1500}$ from the posterior distribution. To approximate this distribution, we fit a finite mixture of multivariate Gaussian's using the penalized likelihood approach of \cite{chen2009inference} in combination with the graphical LASSO \citep{friedman2008sparse} to induce sparsity. A detailed description of this procedure can be found in Section 5 of the SM. The resulting approximation is a mixture of $L=2$ multivariate Gaussian distributions. The $\bm C_\ell$ with respect to each mixture component can be found using \cref{eq:Cmvn} and the total $\bm C$ is found using \cref{eq:Cmixgen}.

\begin{figure}[t]
\centering
\includegraphics[width=\textwidth]{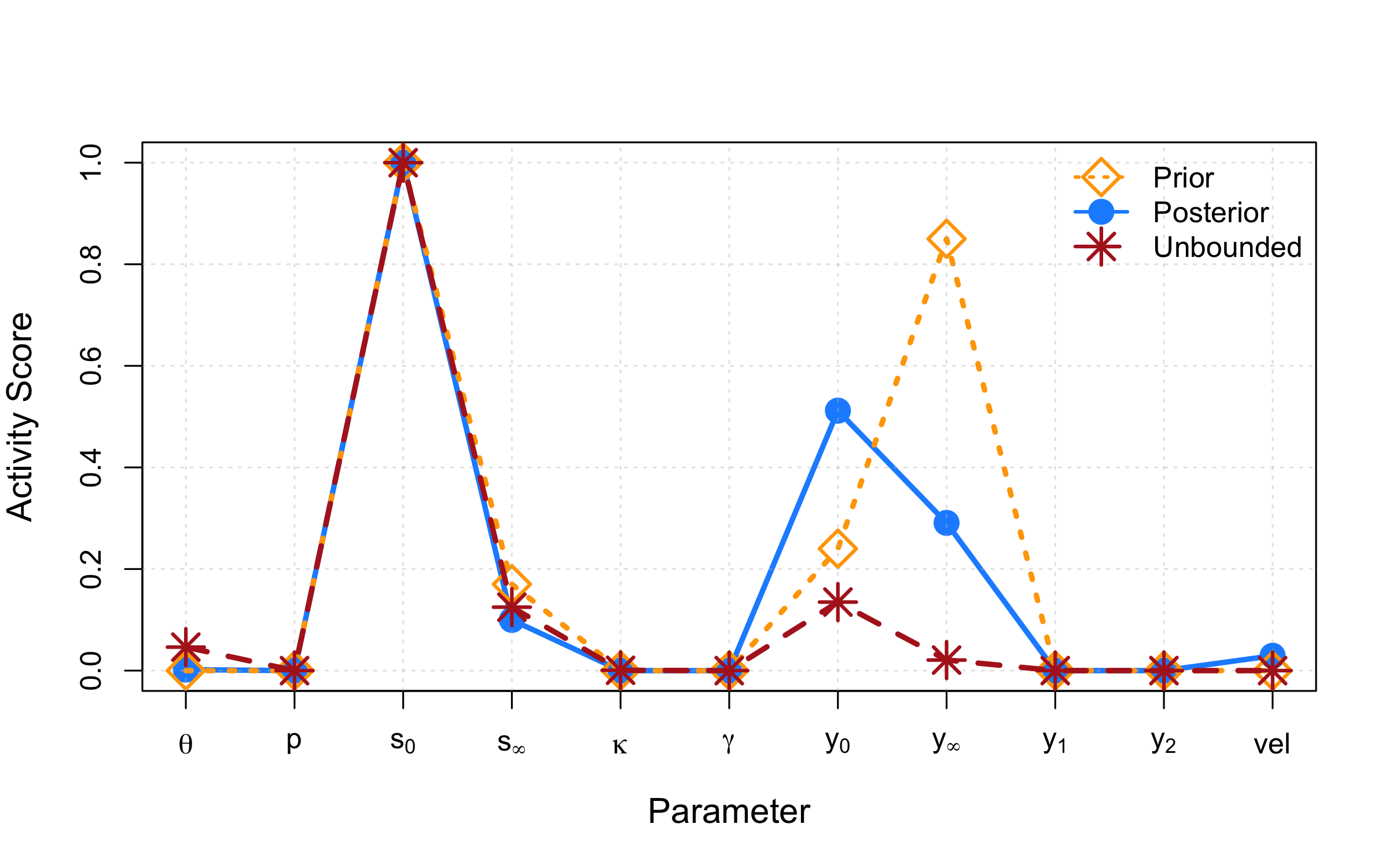}
\caption{Activity scores for the inputs of the PTW Material Strength model with respect to the prior, the calibration posterior and the unbounded space. As the region of parameters space is refined, $y_0$ becomes relatively more important compared to $y_\infty$.  }
\label{fig:activityscores}
\end{figure}

To inspect for any potential differences in the active directions with respect to the prior and the posterior, we examine the activity scores \citep{constantine2017global}, which are defined for the $i^{th}$ input parameter as 
\begin{equation}
    \label{eq:activityscores}
    \text{AS}_i = \sum_{j=1}^p \lambda_j w_{i,j}^2,
\end{equation}
where $\lambda_j$ is the $j^{th}$ eigenvalue and $w_{i,j}$ is the $i^{th}$ entry of the $j^{th}$ eigenvector. Activity scores are a type of global sensitivity metric, with connections to the well-known Sobol index \citep{sobol2001global}. 
For activity score values to be comparable, we use Theorem $1$ to get the $\bm C$ matrix for each variable on a $(0, 1)$ scale before obtaining the eigendecomposition. \cref{fig:activityscores} shows the relative (maximum value of $1.0$) activity scores for each of the PTW input parameters with respect to the prior and the calibration posterior. The two measures lead to similar activity scores, indicating the mostly-stationary behavior of the response across the input space. In both cases, the response variable is only sensitive to four inputs with $s_0$ being the most influential parameter. Across the entire (a priori) input space, the variable $y_\infty$ is the second most influential input. In the a posteriori region of interest, however, the response becomes relatively more sensitive to $y_0$. For comparison, we also include the active subspace, which is obtained by ignoring the constraints in \cref{eq:bounds} altogether, and we see that activity due to $s_0$ is overestimated relative to the other variables and the activity with respect to $y_\infty$ is nearly lost altogether. This analysis shows the sensitivity of active subspace discovery to the specification of the measure and demonstrates the importance of having a method which is compatible with a wide range of appropriate choices for $\rho$.

\subsection{High Dimensional Input: Nuclear Data}
\label{sec:nuclear}
Nuclear data characterize the probability of several types of interactions that can occur between atoms.  For a particular isotope of an element, the probability of each type of interaction is typically a function of energy and is referred to as a cross section. These cross sections are inputs to codes used to simulate various types of nuclear events, such as the behavior of the nuclear material in a nuclear power plant.  Uncertainty in these cross sections, typically parameterized with a multivariate Gaussian distribution, is an important area of research (see chapter 7 of \cite{wu2019neutronics}).

In this example, we consider an analysis for which the nuclear data corresponding to a Plutonium isotope are the inputs to a model, and we want to understand the active subspace.  More specifically, the inputs are eight cross sections, each discretized into 30 energy bins, which results in 240 inputs (135 of which are non-degenerate and linearly independent).  The response is a calculation of criticality ($k_\text{eff}$), for which numbers larger than one indicate the multiplication of neutrons (a chain reaction), while values less than one indicate an unsustained reaction.  The system of interest is a ``critical assembly'' known as Jezebel \citep{sigeti2018endf}, an experiment used to study criticality of Plutonium, as simulated by a neutron transport code.  We have 9372 total simulations, where the 240 dimensional input is generated from the appropriate multivariate Gaussian distribution to reflect underlying uncertainty about the nature of Plutonium (see \cite{sigeti2018endf} for more simulation details).

The purpose of this example is to demonstrate our methodology on a challenging (yet realistic) problem. This problem has high input dimension, many model runs and a highly structured correlated prior. The GP based approach of \cite{wycoff2021sequential} is not at all feasible for this problem and a Monte Carlo based approach will be inefficient given the large input dimension and possibly inaccurate given the lack of well-defined gradients. Prior to this work, the formal active subspace for this computer simulator could not be reasonably and practically obtained, with previous work using the ordinary least squares linear model as a rough proxy \citep{francom2019nuclear}.  Using the methods described in this paper with the Bayesian MARS algorithm of \citep{francom2020bass}, the full active subspace can be discovered using a single core of a 2019 MacBook Pro in under half an hour: $89$ seconds to fit the Bayesian MARS model and $27.47$ minutes to compute the $C$ matrix. 

\begin{figure}[t]
\centering
\includegraphics[width=\textwidth]{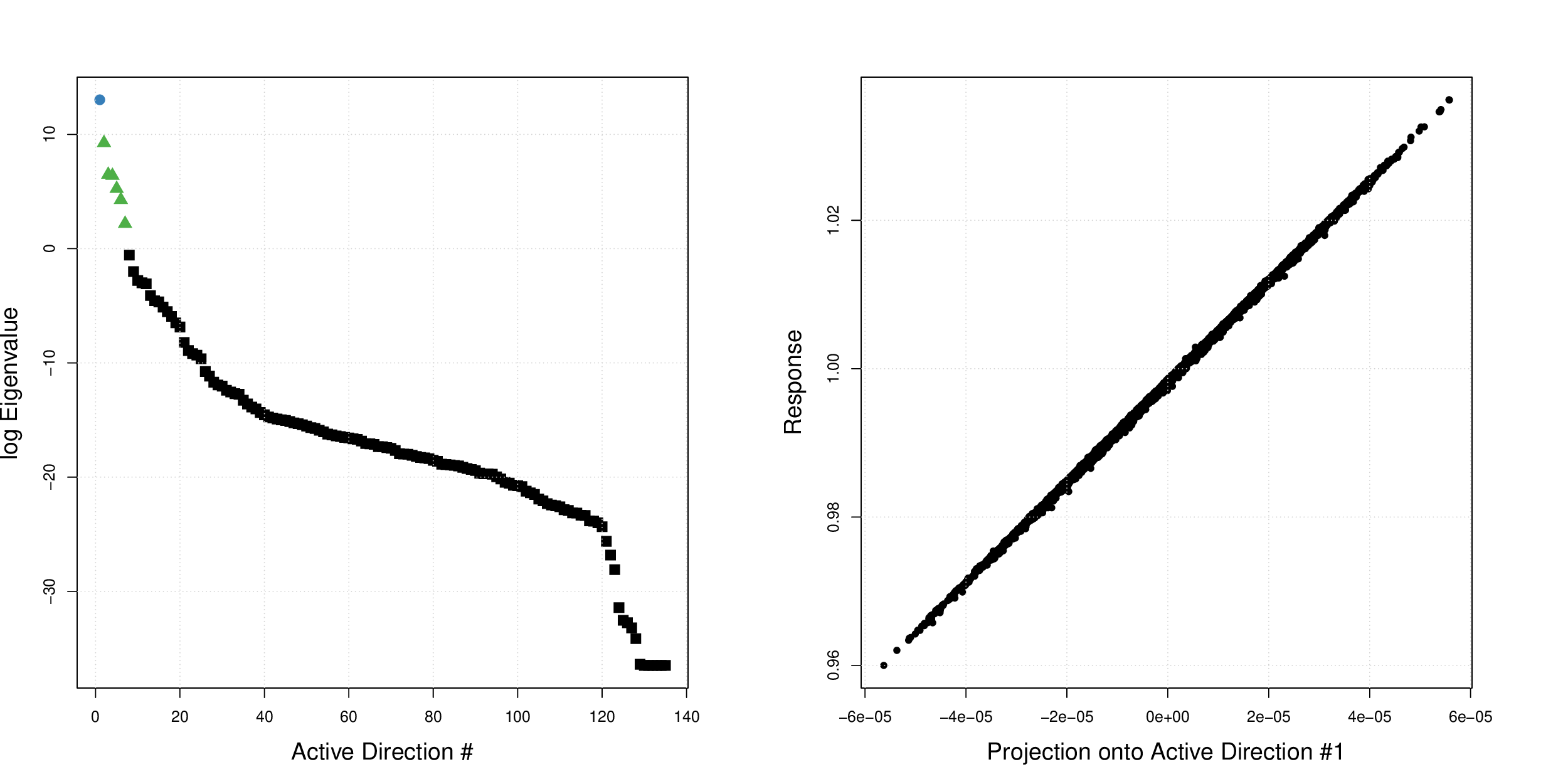}
\caption{The eigenvalues of the estimated $\bm C$ are shown in the left panel. The point symbols indicate that $d=1$ and $d=7$ are reasonable choices for the dimension of the active subspace based on sequential testing. In the right panel, a strong linear relationship can be found between the response and the data projected onto the first active direction.}
\label{fig:nucleardata}
\end{figure}

We briefly summarize the steps required to find the active subspace. Given the prior mean $\bm\mu$ and covariance matrix $\bm\Sigma$, we transform the data as $\bm z = \bm\Sigma^{-1/2}\left(\bm x - \bm \mu\right)$. Since the induced prior for $\bm z$ is now standard normal, we can use \cref{eq:truncmoments} from \cref{sec:univariate} to estimate $\bm C_z$. Finally, we use \cref{thm1} and \cref{eq:Cmvn} to obtain the $\bm C_x$ matrix with respect to the original inputs and obtain the eigendecomposition of this matrix to reveal the active subspace. The left panel of \cref{fig:nucleardata} shows the eigenvalues for the $135$ linearly independent inputs. Using a sequential testing procedure \citep{ma2013review}, $d=1$ and $d=7$ are found to be reasonable choices for the reduced dimension of the active subspace. The right panel of \cref{fig:nucleardata} represents the data, projected onto the first active direction, against the response variable. The fit is exceptionally good, which agrees with previous work which demonstrated the simple linear model adequately emulated the simulator across the input region of interest. 

Finally, we comment on another practical benefit of this procedure. For this computationally intensive dataset, the Bayesian MARS model took $89$ seconds to run, required up to $M=223$ basis functions, and performed no better than a simple linear model. After projecting the original data onto the active directions that were discovered, the Bayesian MARS model which is fit to the projected data took just $38$ seconds to run, required just $M=44$ basis functions (leading to a more memory-efficient model) and outperformed both the linear model and the first BMARS model by an order of magnitude (with respect to RMSE).

\section{Conclusion}
In this work, we discuss the limitations of current methodologies for the discovery of active subspaces for computer models with high dimensional inputs and we propose an alternative approach which is designed for these cases. The new approach is fast and highly accurate, even when a large number of observations are required to capture the full behavior of the computer model. In small sample or small input dimension problems, our approach is still competitive with an existing method based on Gaussian processes \citep{wycoff2021sequential}. Though our development focuses on the use of MARS or Bayesian MARS models as a surrogate, the results in \cref{sec:general} are general enough to recover the GP based solution and can be applied to other surrogate methods including radial basis regression. We have also provided, to date, the most comprehensive discussion of the measure $\rho$ on which the active subspace depends. The new approach is flexible enough to handle univariate measures including beta, gamma and truncated normal distributions, and can also treat more sophisticated measures such as a mixture of multivariate normal distributions. This flexibility handles important special cases, such as prior distributions with physical constraints on the input space and posterior distributions which are represented as a set of finite samples (e.g., calibration). This work can be extended by developing equations for sequential computer experiment design, such as in \cite{wycoff2021sequential}. It is also possible to develop a robustness procedure for the estimation of $C$, by repeating the analysis on the projected data $\bm W^\intercal \bm x$ and using \cref{thm1} until the active subspace converges. This can complicated because the measure corresponding to the projected variables is a high dimensional composition of the measure for $\bm x$ and dependence among the projected variables will be introduced. For similar reasons, producing the relevant calculations for neural network surrogates \citep{tripathy2018deep} represents a highly-challenging yet valuable opportunity for future work. Lastly, as shown in \cref{sec:nuclear}, the active directions can be leveraged to create a new Bayesian MARS model which is both (i) more accurate and (ii) more memory efficient than the model based on the native input space. This suggests that a projected Bayesian MARS algorithm, with similarities to neural networks \citep{mao1995artificial} and Bayesian projection pursuit \citep{collins2022bayesian},  may be a valuable tool for nonparametric regression and computer model emulation.




\bibliographystyle{agsm}
\bibliography{references}

@article{chen2005analytical,
  title={Analytical variance-based global sensitivity analysis in simulation-based design under uncertainty},
  author={Chen, Wei and Jin, Ruichen and Sudjianto, Agus},
  journal={Journal of mechanical design},
  volume={127},
  number={5},
  pages={875--886},
  year={2005}
}

@article{sobol2001global,
  title={{Global sensitivity indices for nonlinear mathematical models and their Monte Carlo estimates}},
  author={Sobol, Ilya M},
  journal={Mathematics and computers in simulation},
  volume={55},
  number={1-3},
  pages={271--280},
  year={2001},
  publisher={Elsevier}
}

@article{preston2003model,
  title={Model of plastic deformation for extreme loading conditions},
  author={Preston, Dean L and Tonks, Davis L and Wallace, Duane C},
  journal={Journal of applied physics},
  volume={93},
  number={1},
  pages={211--220},
  year={2003},
  publisher={American Institute of Physics}
}

@book{wu2019neutronics,
  title={Neutronics of Advanced Nuclear Systems},
  author={Wu, Yican},
  year={2019},
  publisher={Springer}
}

@techreport{sigeti2018endf,
  title={{ENDF/B-VIIIrc1 239Pu Uncertainties: Constraints, Sampling, and Calibration to Jezebel}},
  author={Sigeti, David Edward and Parsons, Donald Kent and White, Morgan Curtis and Francom, Devin Craig and Vander Wiel, Scott Alan and Weaver, Brian Phillip and Williams, Brian J},
  year={2018},
  institution={Los Alamos National Lab.(LANL), Los Alamos, NM (United States)}
}

@article{denison1998bayesian,
  title={{Bayesian MARS}},
  author={Denison, David GT and Mallick, Bani K and Smith, Adrian FM},
  journal={Statistics and Computing},
  volume={8},
  number={4},
  pages={337--346},
  year={1998},
  publisher={Springer}
}

@article{friedman1991,
  title={Multivariate adaptive regression splines},
  author={Friedman, Jerome H},
  journal={The annals of statistics},
  pages={1--67},
  year={1991},
  publisher={JSTOR}
}

@article{tripathy2018deep,
  title={{Deep UQ: Learning deep neural network surrogate models for high dimensional uncertainty quantification}},
  author={Tripathy, Rohit K and Bilionis, Ilias},
  journal={Journal of computational physics},
  volume={375},
  pages={565--588},
  year={2018},
  publisher={Elsevier}
}

@article{francom2020bass,
  title={{BASS: An R package for fitting and performing sensitivity analysis of Bayesian adaptive spline surfaces}},
  author={Francom, Devin and Sans{\'o}, Bruno},
  journal={Journal of Statistical Software},
  volume={94},
  number={1},
  pages={1--36},
  year={2020}
}

@article{wycoff2021sequential,
  title={Sequential Learning of Active Subspaces},
  author={Wycoff, Nathan and Binois, Micka{\"e}l and Wild, Stefan M},
  journal={Journal of Computational and Graphical Statistics},
  volume={30},
  number={4},
  pages={1224--1237},
  year={2021},
  publisher={Taylor \& Francis}
}

@article{binois2021hetgp,
  title={{hetgp: Heteroskedastic Gaussian process modeling and sequential design in R}},
  author={Binois, Micka{\"e}l and Gramacy, Robert B},
  journal={Journal of Statistical Software},
  volume={98},
  pages={1--44},
  year={2021}
}

@techreport{therneau1997introduction,
  title={{An introduction to recursive partitioning using the RPART routines}},
  author={Therneau, Terry M and Atkinson, Elizabeth J and others},
  year={1997},
  institution={Technical report Mayo Foundation}
}

@article{fugate2005hierarchical,
  title={{Hierarchical Bayesian analysis and the Preston-Tonks-Wallace model}},
  author={Fugate, Michael and Williams, Brian and Higdon, David and Hanson, Kenneth M and Gattiker, James and Chen, Shuh-Rong and Unal, Cetin},
  journal={Los Alamos National Laboratory Technical Report LA-UR-05-3935},
  year={2005}
}

@article{kennedy2001bayesian,
  title={Bayesian calibration of computer models},
  author={Kennedy, Marc C and O'Hagan, Anthony},
  journal={Journal of the Royal Statistical Society: Series B (Statistical Methodology)},
  volume={63},
  number={3},
  pages={425--464},
  year={2001},
  publisher={Wiley Online Library}
}

@article{price2013validation,
  title={{Validation of the Preston--Tonks--Wallace strength model at strain rates approaching~ 1011 s- 1 for Al-1100, tantalum and copper using hypervelocity impact crater morphologies}},
  author={Price, Mark C and Kearsley, Anton T and Burchell, Mark J},
  journal={International Journal of Impact Engineering},
  volume={52},
  pages={1--10},
  year={2013},
  publisher={Elsevier}
}

@article{park1994optimalLHS,
  title={{Optimal Latin-hypercube designs for computer experiments}},
  author={Park, Jeong-Soo},
  journal={Journal of statistical planning and inference},
  volume={39},
  number={1},
  pages={95--111},
  year={1994},
  publisher={Elsevier}
}

@article{redner1984mixture,
  title={{Mixture densities, maximum likelihood and the EM algorithm}},
  author={Redner, Richard A and Walker, Homer F},
  journal={SIAM review},
  volume={26},
  number={2},
  pages={195--239},
  year={1984},
  publisher={SIAM}
}

@inproceedings{lukaczyk2014active,
  title={Active subspaces for shape optimization},
  author={Lukaczyk, Trent W and Constantine, Paul and Palacios, Francisco and Alonso, Juan J},
  booktitle={10th AIAA multidisciplinary design optimization conference},
  pages={1171},
  year={2014}
}

@article{ji2019quantifying,
  title={Quantifying kinetic uncertainty in turbulent combustion simulations using active subspaces},
  author={Ji, Weiqi and Ren, Zhuyin and Marzouk, Youssef and Law, Chung K},
  journal={Proceedings of the Combustion Institute},
  volume={37},
  number={2},
  pages={2175--2182},
  year={2019},
  publisher={Elsevier}
}

@article{jefferson2015active,
  title={Active subspaces for sensitivity analysis and dimension reduction of an integrated hydrologic model},
  author={Jefferson, Jennifer L and Gilbert, James M and Constantine, Paul G and Maxwell, Reed M},
  journal={Computers \& geosciences},
  volume={83},
  pages={127--138},
  year={2015},
  publisher={Elsevier}
}

@article{hielscher2018framework,
  title={A framework for expert-driven subpopulation discovery and evaluation using subspace clustering for epidemiological data},
  author={Hielscher, Tommy and Niemann, Uli and Preim, Bernhard and V{\"o}lzke, Henry and Ittermann, Till and Spiliopoulou, Myra},
  journal={Expert Systems with Applications},
  volume={113},
  pages={147--160},
  year={2018},
  publisher={Elsevier}
}

@inproceedings{othmer2016active,
  title={On active subspaces in car aerodynamics},
  author={Othmer, Carsten and Lukaczyk, Trent W and Constantine, Paul and Alonso, Juan J},
  booktitle={17th AIAA/ISSMO Multidisciplinary Analysis and Optimization Conference},
  pages={4294},
  year={2016}
}

@article{constantine2017time,
  title={Time-dependent global sensitivity analysis with active subspaces for a lithium ion battery model},
  author={Constantine, Paul G and Doostan, Alireza},
  journal={Statistical Analysis and Data Mining: The ASA Data Science Journal},
  volume={10},
  number={5},
  pages={243--262},
  year={2017},
  publisher={Wiley Online Library}
}

@article{constantine2015discovering,
  title={Discovering an active subspace in a single-diode solar cell model},
  author={Constantine, Paul G and Zaharatos, Brian and Campanelli, Mark},
  journal={Statistical Analysis and Data Mining: The ASA Data Science Journal},
  volume={8},
  number={5-6},
  pages={264--273},
  year={2015},
  publisher={Wiley Online Library}
}

@inproceedings{batta2021uncovering,
  title={Uncovering Active Structural Subspaces Associated with Changes in Indicators for Alzheimer’s Disease},
  author={Batta, Ishaan and Abrol, Anees and Calhoun, Vince},
  booktitle={2021 43rd Annual International Conference of the IEEE Engineering in Medicine \& Biology Society (EMBC)},
  pages={3948--3951},
  year={2021},
  organization={IEEE}
}

@incollection{tezzele2018combined,
  title={Combined parameter and model reduction of cardiovascular problems by means of active subspaces and POD-Galerkin methods},
  author={Tezzele, Marco and Ballarin, Francesco and Rozza, Gianluigi},
  booktitle={Mathematical and numerical modeling of the cardiovascular system and applications},
  pages={185--207},
  year={2018},
  publisher={Springer}
}

@book{constantine2015activebook,
  title={Active subspaces: Emerging ideas for dimension reduction in parameter studies},
  author={Constantine, Paul G},
  year={2015},
  publisher={SIAM}
}

@article{constantine2016accelerating,
  title={{Accelerating Markov chain Monte Carlo with active subspaces}},
  author={Constantine, Paul G and Kent, Carson and Bui-Thanh, Tan},
  journal={SIAM Journal on Scientific Computing},
  volume={38},
  number={5},
  pages={A2779--A2805},
  year={2016},
  publisher={SIAM}
}

@article{seshadri2018turbomachinery,
  title={Turbomachinery active subspace performance maps},
  author={Seshadri, Pranay and Shahpar, Shahrokh and Constantine, Paul and Parks, Geoffrey and Adams, Mike},
  journal={Journal of Turbomachinery},
  volume={140},
  number={4},
  year={2018},
  publisher={American Society of Mechanical Engineers Digital Collection}
}

@article{constantine2017global,
  title={Global sensitivity metrics from active subspaces},
  author={Constantine, Paul G and Diaz, Paul},
  journal={Reliability Engineering \& System Safety},
  volume={162},
  pages={1--13},
  year={2017},
  publisher={Elsevier}
}

@article{constantine2014activekriging,
  title={Active subspace methods in theory and practice: applications to kriging surfaces},
  author={Constantine, Paul G and Dow, Eric and Wang, Qiqi},
  journal={SIAM Journal on Scientific Computing},
  volume={36},
  number={4},
  pages={A1500--A1524},
  year={2014},
  publisher={SIAM}
}

@article{hardle1989investigating,
  title={Investigating smooth multiple regression by the method of average derivatives},
  author={H{\"a}rdle, Wolfgang and Stoker, Thomas M},
  journal={Journal of the American statistical Association},
  volume={84},
  number={408},
  pages={986--995},
  year={1989},
  publisher={Taylor \& Francis}
}

@article{friedman1981projection,
  title={Projection pursuit regression},
  author={Friedman, Jerome H and Stuetzle, Werner},
  journal={Journal of the American statistical Association},
  volume={76},
  number={376},
  pages={817--823},
  year={1981},
  publisher={Taylor \& Francis}
}

@article{xia2008multiple,
  title={A multiple-index model and dimension reduction},
  author={Xia, Yingcun},
  journal={Journal of the American Statistical Association},
  volume={103},
  number={484},
  pages={1631--1640},
  year={2008},
  publisher={Taylor \& Francis}
}

@article{li1991sliced,
  title={Sliced inverse regression for dimension reduction},
  author={Li, Ker-Chau},
  journal={Journal of the American Statistical Association},
  volume={86},
  number={414},
  pages={316--327},
  year={1991},
  publisher={Taylor \& Francis}
}

@article{friedman2008sparse,
  title={Sparse inverse covariance estimation with the graphical lasso},
  author={Friedman, Jerome and Hastie, Trevor and Tibshirani, Robert},
  journal={Biostatistics},
  volume={9},
  number={3},
  pages={432--441},
  year={2008},
  publisher={Oxford University Press}
}

@article{chen2009inference,
  title={Inference for multivariate normal mixtures},
  author={Chen, Jiahua and Tan, Xianming},
  journal={Journal of Multivariate Analysis},
  volume={100},
  number={7},
  pages={1367--1383},
  year={2009},
  publisher={Elsevier}
}

@techreport{francom2019nuclear,
  title={Nuclear Data Dimension Reduction},
  author={Francom, Devin Craig and Vander Wiel, Scott Alan and Weaver, Brian Phillip},
  year={2019},
  institution={Los Alamos National Lab.(LANL), Los Alamos, NM (United States)}
}

@inproceedings{bierens1987kernel,
  title={Kernel estimators of regression functions},
  author={Bierens, Herman J},
  booktitle={Advances in econometrics: Fifth world congress},
  volume={1},
  pages={99--144},
  year={1987}
}

@incollection{higdon2002space,
  title={Space and space-time modeling using process convolutions},
  author={Higdon, Dave},
  booktitle={Quantitative methods for current environmental issues},
  pages={37--56},
  year={2002},
  publisher={Springer}
}

@article{rumsey2023generalized,
  title={Generalized Bayesian MARS: Tools for Emulating Stochastic Computer Models},
  author={Rumsey, Kellin and Francom, Devin and Shen, Andy},
  journal={arXiv preprint arXiv:2306.01911},
  year={2023}
}

@article{friedman2003multiple,
  title={Multiple additive regression trees with application in epidemiology},
  author={Friedman, Jerome H and Meulman, Jacqueline J},
  journal={Statistics in medicine},
  volume={22},
  number={9},
  pages={1365--1381},
  year={2003},
  publisher={Wiley Online Library}
}

@book{gramacy2020surrogates,
  title={Surrogates: Gaussian process modeling, design, and optimization for the applied sciences},
  author={Gramacy, Robert B},
  year={2020},
  publisher={CRC press}
}

@article{datta2014bootstrap,
  title={Bootstrap—an exploration},
  author={Datta, Jyotishka and Ghosh, Jayanta K},
  journal={Statistical Methodology},
  volume={20},
  pages={63--72},
  year={2014},
  publisher={Elsevier}
}

@article{collins2022bayesian,
  title={Bayesian Projection Pursuit Regression},
  author={Collins, Gavin and Francom, Devin and Rumsey, Kellin},
  journal={arXiv preprint arXiv:2210.09181},
  year={2022}
}

@article{mao1995artificial,
  title={Artificial neural networks for feature extraction and multivariate data projection},
  author={Mao, Jianchang and Jain, Anil K},
  journal={IEEE transactions on neural networks},
  volume={6},
  number={2},
  pages={296--317},
  year={1995},
  publisher={IEEE}
}

@article{claeskens2008model,
  title={Model selection and model averaging},
  author={Claeskens, Gerda and Hjort, Nils Lid and others},
  journal={Cambridge Books},
  year={2008},
  publisher={Cambridge University Press}
}

@article{domingos2012few,
  title={A few useful things to know about machine learning},
  author={Domingos, Pedro},
  journal={Communications of the ACM},
  volume={55},
  number={10},
  pages={78--87},
  year={2012},
  publisher={ACM New York, NY, USA}
}

@book{mishra2012algorithmic,
  title={Algorithmic algebra},
  author={Mishra, Bhubaneswar},
  year={2012},
  publisher={Springer Science \& Business Media}
}

@article{ma2013review,
  title={A review on dimension reduction},
  author={Ma, Yanyuan and Zhu, Liping},
  journal={International Statistical Review},
  volume={81},
  number={1},
  pages={134--150},
  year={2013},
  publisher={Wiley Online Library}
}

@article{berger1990robust,
  title={{Robust Bayesian analysis: sensitivity to the prior}},
  author={Berger, James O},
  journal={Journal of statistical planning and inference},
  volume={25},
  number={3},
  pages={303--328},
  year={1990},
  publisher={Elsevier}
}

@misc{truncNorm,
  author = {Fernandez-de-Cossio-Diaz, Jorge},
  title = {TruncatedNormal.jl},
  year = {2017},
  publisher = {GitHub},
  journal = {GitHub repository},
  howpublished = {\url{https://github.com/cossio/TruncatedNormal.jl}}
 }

@article{xu2009adaptive,
  title={Adaptive hinging hyperplanes and its applications in dynamic system identification},
  author={Xu, Jun and Huang, Xiaolin and Wang, Shuning},
  journal={Automatica},
  volume={45},
  number={10},
  pages={2325--2332},
  year={2009},
  publisher={Elsevier}
}

@article{lin1992canonical,
  title={Canonical piecewise-linear approximations},
  author={Lin, J-N and Unbehauen, Rolf},
  journal={IEEE Transactions on Circuits and Systems I: Fundamental Theory and Applications},
  volume={39},
  number={8},
  pages={697--699},
  year={1992},
  publisher={IEEE}
}

@article{shekhtman1982piecewise,
  title={Why Piecewise Linear Functions Are Dense in CIO, 1},
  author={Shekhtman, Boris},
  journal={Journal of Approximation Theory},
  volume={36},
  pages={265--267},
  year={1982}
}

@article{park1993approximation,
  title={Approximation and radial-basis-function networks},
  author={Park, Jooyoung and Sandberg, Irwin W},
  journal={Neural computation},
  volume={5},
  number={2},
  pages={305--316},
  year={1993},
  publisher={MIT Press One Rogers Street, Cambridge, MA 02142-1209, USA journals-info~…}
}

@Manual{activegp,
    title = {activegp: Gaussian Process Based Design and Analysis for the Active
Subspace Method},
    author = {Nathan Wycoff and Mickael Binois},
    year = {2021},
    note = {R package version 1.0.6},
    url = {https://CRAN.R-project.org/package=activegp},
  }

@article{vohra2019active,
  title={Active subspace-based dimension reduction for chemical kinetics applications with epistemic uncertainty},
  author={Vohra, Manav and Alexanderian, Alen and Guy, Hayley and Mahadevan, Sankaran},
  journal={Combustion and Flame},
  volume={204},
  pages={152--161},
  year={2019},
  publisher={Elsevier}
}

@article{navaneeth2022surrogate,
  title={Surrogate assisted active subspace and active subspace assisted surrogate—A new paradigm for high dimensional structural reliability analysis},
  author={Navaneeth, N and Chakraborty, Souvik},
  journal={Computer Methods in Applied Mechanics and Engineering},
  volume={389},
  pages={114374},
  year={2022},
  publisher={Elsevier}
}

\pagebreak
\begin{center}
\textbf{\large Supplemental Materials: Discovering Active Subspaces for High Dimensional Computer Models}
\end{center}
\setcounter{equation}{0}
\setcounter{figure}{0}
\setcounter{table}{0}
\setcounter{page}{1}
\makeatletter
\renewcommand{\theequation}{SM\arabic{equation}}
\renewcommand{\thefigure}{SM\arabic{figure}}
\renewcommand{\bibnumfmt}[1]{[SM#1]}
\renewcommand{\citenumfont}[1]{SM#1}

\section{Estimate C - Pseudocode}

\begin{algorithm}[H]
\caption{Estimate C}
\label{alg:alg1}
\begin{algorithmic}[1]
\onehalfspacing
\Require Data: $\{(\bm x_i, y_i)\}_{i=1}^n$, Measure: $\rho_1, \ldots, \rho_p$
\State \texttt{Model} $\leftarrow$ \texttt{Train-MARS}$(\{(\bm x_i, y_i)\}_{i=1}^n)$ \Comment{e.g., BASS R package}
\For{$i=1$ to $p$}
\For{$m_1=1$ to \texttt{Model}.$M$}
\For{$m_2=1$ to \texttt{Model}.$M$}
\If{$m_1 \leq m_2$} \Comment{Symmetric matrices}
\State \texttt{Compute} $a^{(i)}[m_1, m_2]$ using \texttt{Model} \Comment{Eq. (15)} 
\State \texttt{Compute} $b^{(i)}[m_1, m_2]$ using \texttt{Model}  \Comment{Eq. (15)}
\State \texttt{Compute} $I_2^{(i)}[m_1, m_2]$ using \texttt{Model}, $a$, $b$ and $\rho_i$  \Comment{Eq. (18)} 
\State \texttt{Compute} $I_2^{(i)}[m_1, m_2]$ using \texttt{Model}, $a$, $b$ and $\rho_i$  \Comment{Eq. (19)} 
\EndIf
\State \texttt{Compute} $I_1^{(i)}[m_1, m_2]$ using \texttt{Model}, $a$, $b$ and $\rho_i$  \Comment{Eq. (17)} 
\EndFor
\EndFor
\EndFor
\For{$i=1$ to $p$}
\For{$j=1$ to $p$}
\If{$i == j$}
\State \texttt{Compute} $\bm C_{ii}$ using \texttt{Model}, $I_3^{(i)}$ and $I_2$ \Comment{Eq. (9)}
\Else
\State \texttt{Compute} $\bm C_{ij}$ using \texttt{Model}, $I_1^{(i)}$, $I_1^{(j)}$ and $I_2$ \Comment{Eq. (9)}
\EndIf
\EndFor
\EndFor
\Return $\bm C$
\end{algorithmic}
\end{algorithm}

\section{Proof of Theorem 1}

\begin{theorem}
\label{thm1}
Let $\bm z = \bm A\bm x + \bm b$ and take $\tilde f(\bm z) = f(\bm x)$. If $\bm A$ is invertible, then we have
$$\bm C = \bm A^\intercal \tilde{\bm C} \bm A$$
where $\bm C$ is the $C$ matrix taken with respect to $f$ and $\rho_x$ and $\tilde{ \bm C}$ is taken with respect to $\tilde f$ and $\rho_z(z) = \rho_x(\bm A^{-1}(\bm z - \bm b))/ |\bm A|$.
\end{theorem}

\noindent {\bf Proof:}

\noindent Let $\tilde f (\bm z)$ be a function of the form shown in Equation (6) of the main text and take $f(\bm x) = \tilde f(\bm z (\bm x))$.  Then 
\begin{align*}
    \bm \nabla^T f(\bm x) & =
    \bm \nabla^T \tilde f(\bm z)
    \left(\frac{\partial \bm z}{\partial \bm x^T}\right) \\
    & = \bm \nabla^T \tilde f(\bm z) \bm A, \ \ \text{so that} \\ 
    \bm \nabla f(\bm x) 
    & = \bm A^T \bm \nabla \tilde f(\bm z) .
\end{align*}

Therefore, the $C$ matrix for $f$ can be written in terms of the $C$ matrix for $\tilde f$ as follows. 
\begin{align*}
   \bm C_f & = 
    \int \bm \nabla f(\bm x) \bm \nabla^T f(\bm x) \cdot p_x(\bm x) \; d \bm x \\
    & =
    \int
    \bm A^T \bm \nabla \tilde f(\bm z (\bm x)) \bm \nabla^T \tilde f(\bm z (\bm x))  \bm A \cdot p_x(\bm x) \; d \bm x  \\
    & =
    \bm A^T \left(
    \int \bm \nabla \tilde f(\bm z ) \bm \nabla^T \tilde f(\bm z ) \cdot p_x(\bm A^{-1} (\bm z - \bm b)) \left| \frac{\partial \bm x}{\partial \bm z^T}\right| \; d \bm z
    \right) \bm A \\
    & =
    \bm A^T \left(
    \int \bm \nabla \tilde f(\bm z ) \bm \nabla^T \tilde f(\bm z ) \cdot p_x(\bm A^{-1} (\bm z - \bm b)) \left|\bm A^{-1}\right| \; d \bm z
    \right) \bm A \\
    &= \bm A^T \bm C_{\tilde f} \bm A 
\end{align*}

\noindent $\square$

\begin{figure}[h]
\centering
\includegraphics[width=0.6\textwidth]{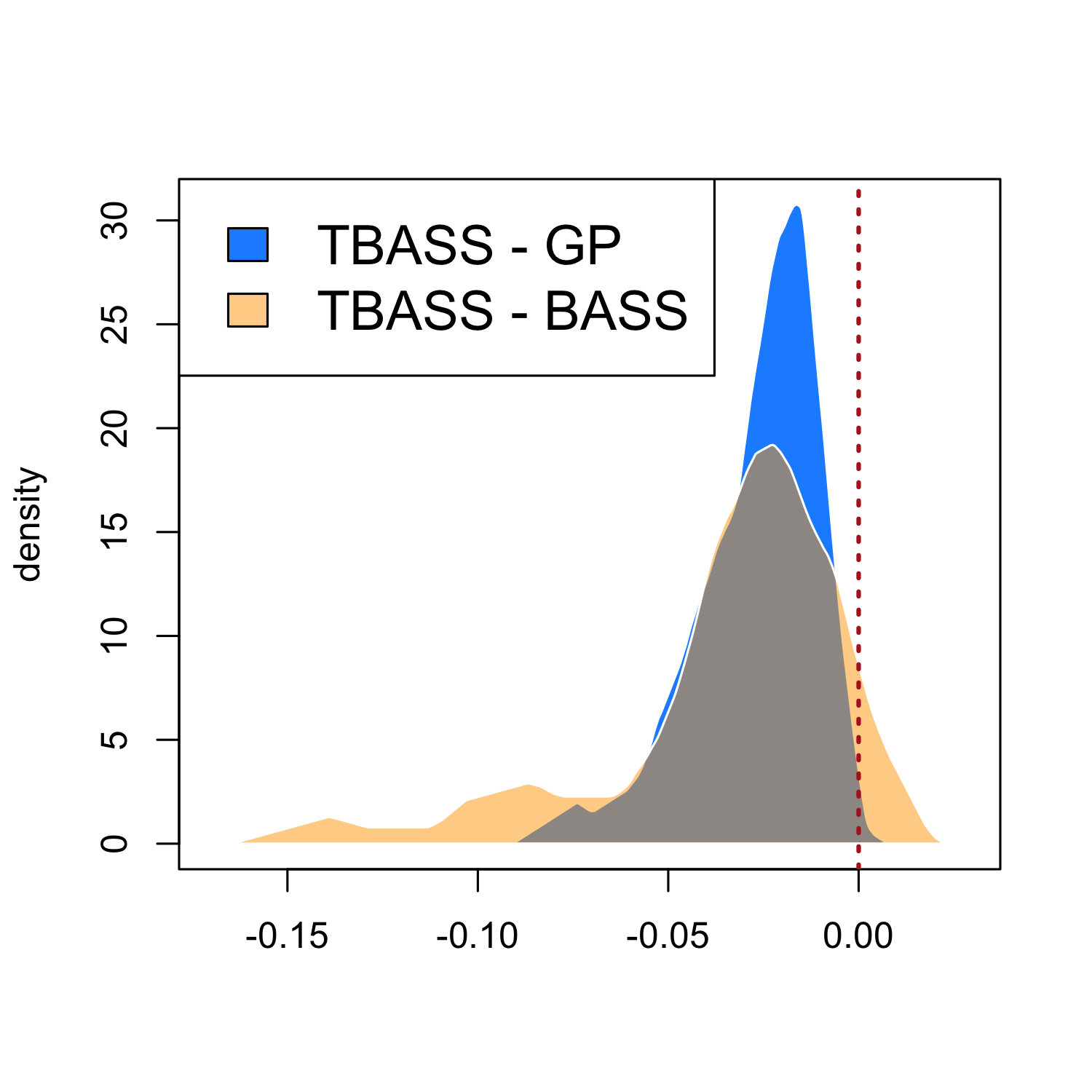}
\caption{Kernel density estimates for the difference in Frobenius error between TBASS and the other two methods across the $30$ simulations. The densities indicate that TBASS had the lowest error in nearly all $30$ simulations. }
\label{fig:poly_outliers}
\end{figure}

\section{Additional Examples}
\subsection{Simple Polynomial with Corrupted Data}
Consider again the simple polynomial function from Section 4.1
$$f(\bm x) = x_1^2 + x_1x_2 + x_2^3/9,$$
and suppose that we are given training data
\begin{align*}
    y_i &= f(\bm x_i) + \xi_i, i=1,2,\ldots, 500 \\
    \xi_i &\sim \begin{dcases}
    N(0, 0.5^2), & i \leq 10 \\
    0, & i > 0.
    \end{dcases}
\end{align*}
Here, $\xi_i$ represents a ``corruption mechanism" which has perturbed the response for $2\%$ of the $n=500$ observations. In this example, we seek to explore the impact of these outliers on our ability to estimate $C$ across various methods. We will show that estimates of $C$ based on GPs \citep{wycoff2021sequential} and those based on standard Gaussian BMARS \citep{francom2020bass} will be negatively affected by this process. Using a Bayesian MARS model under a $t$-likelihood, as explored in \cite{rumsey2023generalized}, can lead to better estimates of $f$ and will result in better estimates of the active subspace. This generalized BMARS approach is compatible with the methods described in this manuscript without requiring any modifications. Code to reproduce this example can be found at \url{https://github.com/knrumsey/ASM-BMARS-Examples}. 

Setting $p =12$, we simulate $30$ different sets of training data using the process described above and we estimate $C$ using a GP, BASS and TBASS($\nu=5$) surrogate. The quality of the estimate of $C$ is measured with the Frobenius norm, and the results are given in \cref{tab:outliers}. The TBASS method is the clear winner, by about an order of magnitude. \Cref{fig:poly_outliers} gives kernel density estimates for the difference in Frobenius norm across the $30$ simulations. 

\begin{table}[!htbp] \centering 
  \caption{Summary of results (Frobenius error) for the simple polynomial function with outliers.} 
  \label{tab:outliers} 
\begin{tabular}{@{\extracolsep{5pt}} lccc} 
\\[-1.8ex]\hline 
\hline \\[-1.8ex] 
 & BASS & TBASS & GP \\ 
\hline \\[-1.8ex] 
Min. & $0.004$ & $0.0003$ & $0.010$ \\ 
1st Qu. & $0.019$ & $0.002$ & $0.017$ \\ 
Median & $0.030$ & $0.002$ & $0.024$ \\ 
Mean & $0.037$ & $0.002$ & $0.029$ \\ 
3rd Qu. & $0.041$ & $0.003$ & $0.037$ \\ 
Max. & $0.142$ & $0.004$ & $0.077$ \\ 
\hline \\[-1.8ex] 
\end{tabular} 
\end{table} 

\subsection{Additional Example: Global Sensitivity Analysis for Stochastic SIR Model}

By estimating $C$ conditional on a MARS model, we are able to take advantage of the flexible-likelihoods offered by generalized Bayesian MARS \citep{rumsey2023generalized}. In this example, we will recreate the example found in Section 4.3 of \cite{rumsey2023generalized} for activity scores (rather than Sobol effects, as in the cited paper). For a detailed description of the problem, we defer to \cite{rumsey2023generalized}. 

\begin{figure}[h]
\centering
\includegraphics[width=0.6\textwidth]{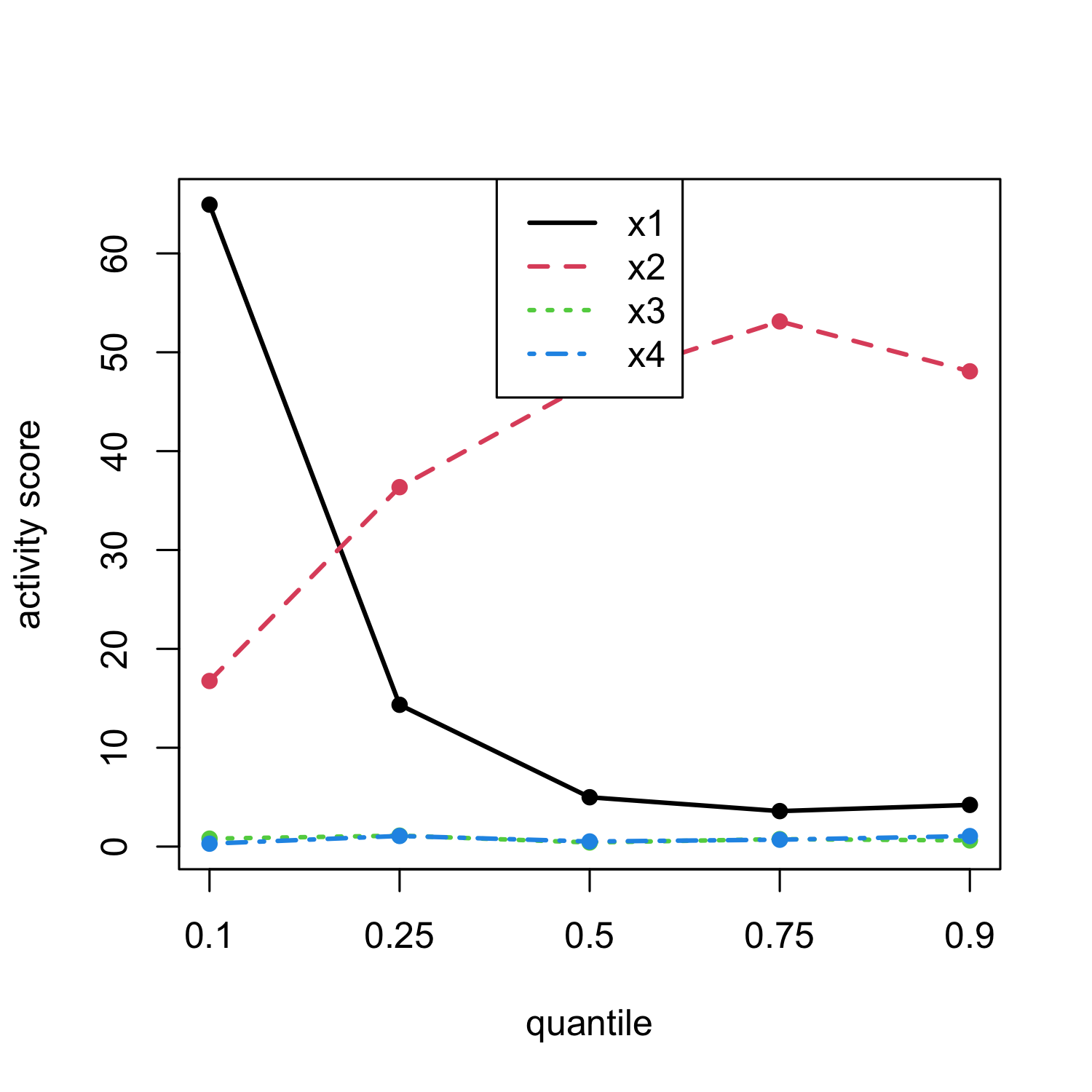}
\caption{Activity scores as a function of quantile for the stochastic SIR model described in Section 4.3 of \cite{rumsey2023generalized}. }
\label{fig:active_SIR}
\end{figure}

Briefly, we take $f$ to be a stochastic SIR model with $p=5$ inputs and we fit a series of quantile regression BASS models for various quantiles $q \in \{0.1, 0.25, 0.50, 0.75, 0.90\}$. We can now assess the sensitivity using activity scores \citep{constantine2015activebook} for each quantile. Much can be learned about a computer model by studying how the sensitivity to its various inputs changes as a function of quantiles. The results of this simulation can be seen in \cref{fig:active_SIR}. Code to reproduce this example can be found at \url{https://github.com/knrumsey/ASM-BMARS-Examples}. 

\subsection{Uncertainty Quantification for Activity Scores}

\begin{figure}[h]
\centering
\includegraphics[width=0.9\textwidth]{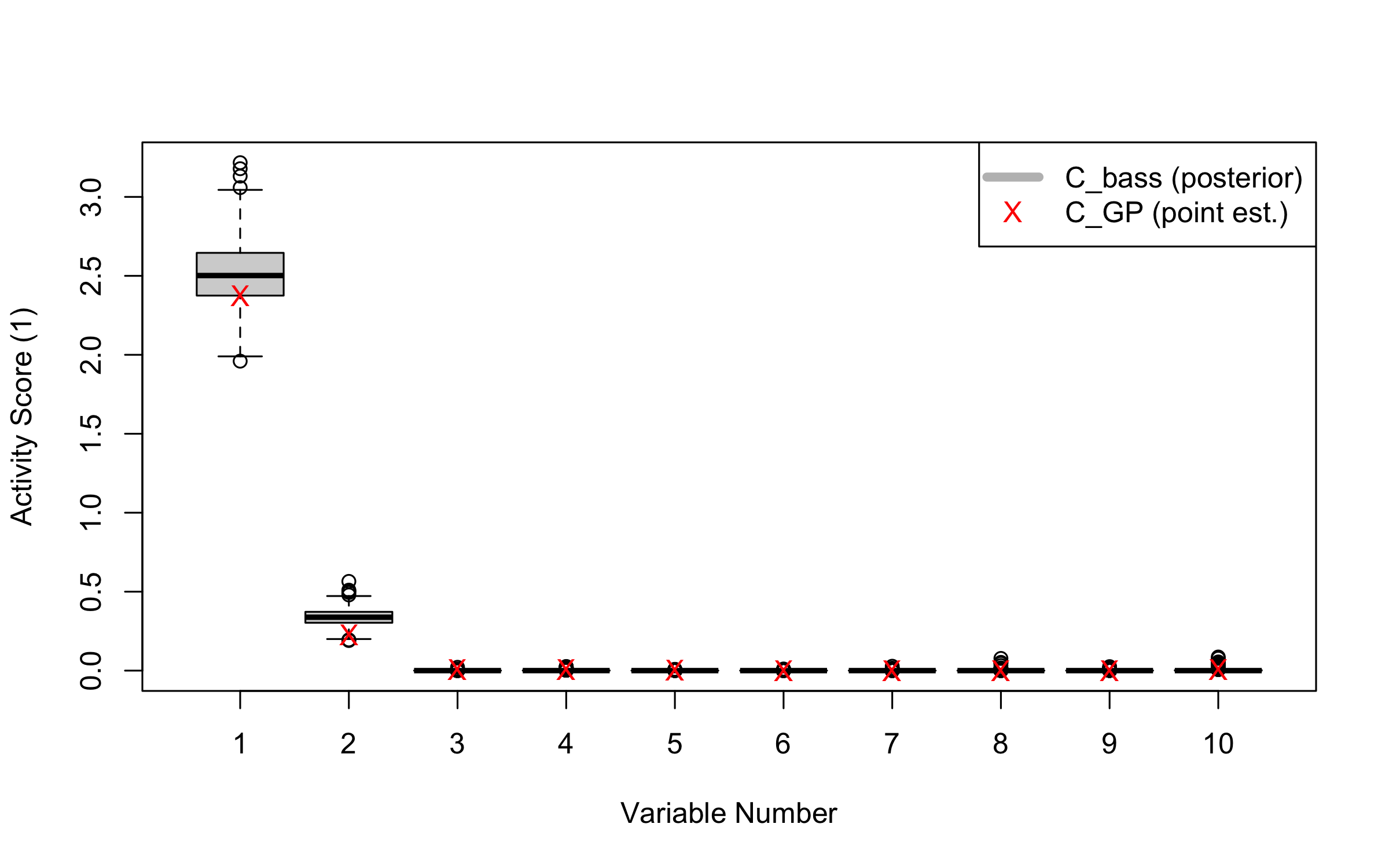}
\caption{Posterior distributions of the activity scores for the simple polynomial model. }
\label{fig:activity_post}
\end{figure}

One advantage of the methods described in this paper, compared to the GP-based approach, is that the posterior uncertainty of the surrogate model fit can be propagated forward to the estimate of the active subspace. In this example, we return again to the simple polynomial function of Section 4.1. With $n=100$ and $p=10$, we fit a BASS model which can be viewed as a posterior ensemble of various MARS models, one for each MCMC iteration. We can find the $C$ matrix corresponding to each MARS model. By computing the activity scores \citep{constantine2015activebook} for each $C$, we obtain posterior distributions for these quantities. \Cref{fig:activity_post} shows the posterior distribution of the activity scores (as boxplots) alongside the point estimate produced by the GP-based approach. Code to reproduce this example can be found at \url{https://github.com/knrumsey/ASM-BMARS-Examples}. 

\section{Details of recursive partitioning algorithm}

 We begin by finding the minimal rectangle which contains $\mathcal X$, call it $R_0$, which is fully defined by the $(a_{k0}, b_{k0})$ spanning each of the $k=1,\ldots,d$ coordinates. We then split $R_0$ along one of it's dimensions $k$ forming two new boxes $R_1$ and $R_2$ such that 
$$a_{k1} = a_{k0} \quad\quad b_{k1} = \frac{a_{k0} + b_{k0}}{2} \quad\quad a_{k2} = \frac{a_{k0} + b_{k0}}{2} \quad\quad b_{k2} = b_{k0}$$
and $a_{j0} = a_{j1} = a_{j2}$, $b_{j0} = b_{j1} = b_{j2}$ for $j \neq k$, where $(a_{jc}, b_{jc})$. We then check (separately) to see if $R_1$ and $R_2$ is either (i) entirely contained in $\mathcal X$, (ii) entirely contained outside of $\mathcal X$ or (iii) partially contained in $\mathcal X$. In the first case, we terminate and return the box. In the second case, we terminate without returning anything and in the final case we recursively repeat the procedure, treating the box $R_1$ (or $R_2$) as a new $R_0$ and using a new splitting dimension $k'$. 

This procedure requires that we can quickly check whether a box $R$ lives inside, outside or overlaps the region $\mathcal X$. If $\mathcal X$ is concave, as is it is in the PTW model, then it is sufficient to check if each of the $2^p$ vertices is in $\mathcal X$ (with at least one vertex required to be in the interior of $\mathcal X$, for $R$ to be considered entirely contained). There is also some freedom in the mechanism for choosing the splitting dimension $k$. For instance, in the PTW example, we only split over variables $x_3, x_4, x_7, x_8$ and $x_9$ and we iterate between these dimensions at each level of recursion (similarly, we need only check the $2^5$ vertices corresponding to these inputs, rather than the full $2^{11}$). Finally, we note that this procedure will converge to $\mathcal X$ (under Lesbesgue measure) for regions $\mathcal X$ which can be defined by a set of linear inequalities \citep{mishra2012algorithmic}, and so a reasonable stopping criteria is needed. One sensible choice is to stop recursing when the rectangle volume is below a pre-specified threshold.

Using this procedure for the PTW model with the bounds given in Table 2 (main text), the constraint in Equation 35 (main text) and a minimum box volume of $10^{-12}$, we were able to generate an approximation consisting of $L = 478$ disjoint boxes. These boxes cover only about $54\%$ of the constrained space by volume, but volume is a highly misleading measure in high dimensions. This is an unsurprising result, since volume in high dimensions is concentrated around the boundary ("nearly all the volume in a high dimensional orange is in the peel" \citep{domingos2012few}) and the recursive method described here fails to fill in edges of the space, because the boxes required for "squeezing" into corners are smaller than the tolerance. By lowering the tolerance to $10^{-13}$, we obtain an approximation consisting of $11,920$ boxes which covers $84\%$ of the space by volume. 

\section{Details of multivariate normal mixture fitting}

For the PTW example, we are given $1500$ draws from the $11$-dimensional posterior distribution. Since fitting mixtures of multivariate Gaussian distributions is a computationally demanding and difficult problem, we start by reducing the dimension of the mixture from $11$ to just the $5$ dimensions that are involved with the PTW constraint in Equation 35. This choice was supported via the graphical LASSO approach of \cite{friedman2008sparse} and the bootstrap as applied in Section 2 of \cite{datta2014bootstrap}. Both approaches suggest that the $6$ remaining inputs (those not included in the constraint) can be treated as independent.

\begin{figure}[t]
\centering
\includegraphics[width=\textwidth]{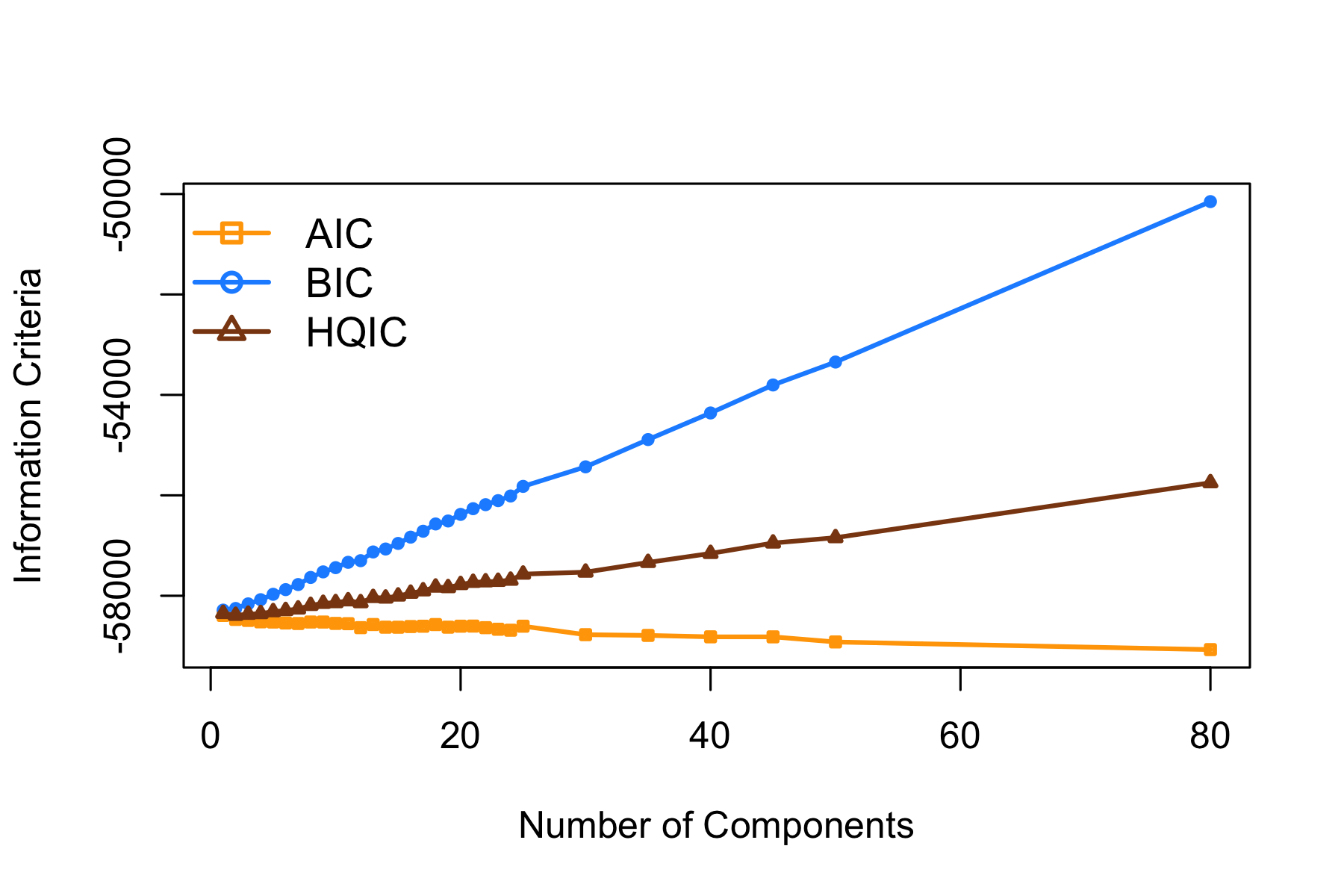}
\caption{Model selection criterion for mixture of multivariate Gaussians fit to posterior draws from PTW model. In repeated experiments, HQIC prefers two mixture components.  }
\label{fig:activityscores}
\end{figure}

From here we apply the robust penalized EM algorithm of \cite{chen2009inference}, which shrinks the MLE of the covariance matrix towards the sample covariance matrix. We use a small penalty term for robustness, corresponding to $n^{-1}$ ($n=1500$) in \cite{chen2009inference}. The log-likelihood grows roughly linearly with the number of components (from $L=1$ to $L=80$). The rate of growth is such that the Bayesian information criterion (BIC) suggests a standard multivariate Gaussian is appropriate and Akaike's information criterion (AIC) suggests $L > 50$ mixture components. On the other hand, the Hannan-Quinn information criterion (HQIC) repeatedly (running multiple experiments to reduce convergence error) prefers $L=2$ mixture components, although it largely agrees with BIC. \cite{claeskens2008model} discuss the advantages of HQIC, which include strong consistency (unlike AIC) and near asymptotic efficiency (unlike BIC, missing optimal convergence by a very small $\log\log n$ factor). Monte Carlo suggests that this 2-component mixture of multivariate Gaussians has about $96\%$ of it's density inside the constraints of Equation 15 (main text). The two component mixture that we select for inputs $(s_0, s_\infty, y_0, y_\infty, y_1)$ is given by its parameters 
\begin{equation}
\begin{aligned}
    \pi &= (0.701, 0.299) \\
    \mu_1 &= [0.031 \ 0.015 \ 0.020 \ 0.005 \ 0.067]^\intercal \\
    \mu_2 &= [0.028 \ 0.010 \ 0.020 \ 0.006 \ 0.064 ]^\intercal \\
    \Sigma_1 &= 5869.309 \begin{bmatrix}
    0.22346 & 0.01801  & 0.08197  & 0.00293  & 0.07142 \\
    0.01801 & 0.13125  & 0.00410  & 0.02480  & 0.00387 \\
    0.08197 & 0.00410  & 0.13088 & -0.00127 & -0.00091 \\
    0.00293 & 0.02480 & -0.00127  & 0.02376  & 0.00206 \\
    0.07142 & 0.00387 & -0.00091  & 0.00206  & 1.00000 \\
    \end{bmatrix} \\
    \Sigma_2 &= 5866.06 \begin{bmatrix}
0.18100 & -0.00557  & 0.10496 & -0.01146 & 0.06603 \\
-0.00557  & 0.05328  & 0.00203  & 0.02379 & 0.02637 \\
0.10496  & 0.00203  & 0.13695 & -0.00439 & 0.04153 \\
-0.01146  & 0.02379 & -0.00439  & 0.02314 & 0.01042 \\
0.06603  & 0.02637  & 0.04153  & 0.01042 & 1.00000 
    \end{bmatrix},
\end{aligned}
\end{equation}
and the remaining inputs are fitted with independent univariate normal distributions.

\section{Propagation of Model Uncertainty with Bayesian MARS}
\label{sec:posterior}


For all of the results in this paper, we construct surrogate models using the Bayesian MARS algorithm of \citep{francom2020bass}. One advantage of Bayesian MARS is that the posterior distribution of the model parameters essentially yields an ensemble of MARS models. By computing the active subspace for each model in the ensemble, we can (i) improve the final estimate of the active subspace and (ii) characterize the uncertainty in the estimate that comes from the model fitting procedure. This comes at a cost of course, since the $\bm C$ matrix must be constructed $G$ times, where $G$ is the number of samples from the posterior distribution. For large complex problems, constructing $\bm C$ may be fast, but if $G$ is large, the ensemble can take a long time to process. 

In order to leverage the structure of the Bayesian MARS posterior sampler, we first write 
\begin{equation}
\label{eq:Cijv2}
    \bm C_{ij} = 
    \begin{dcases}
    \bm\gamma\left(I_1^{(i)} \odot {I_1^{(j)}}^\intercal \bigodot_{k\not\in\{i,j\}}I_2^{(k)}\right)\bm\gamma^\intercal, & i = j \\[1.5ex]
    \bm\gamma\left(I_3^{(i)} \bigodot_{k\neq i}I_2^{(k)}\right)\bm\gamma^\intercal, & i \neq j 
    \end{dcases},
\end{equation}
where $\odot$ represents the (elementwise) Hadamard product. This formula provides a convenient way to compute each element of $\bm C$ quickly, and in terms of the matrices $I_1^{(i)}, I_2^{(i)}$ and $I_3^{(i)}$.

Through close examination of the sampling procedure used in Bayesian MARS, we can find opportunities for substantial time savings. Using $\mathcal O(pM^2)$ memory, we can reuse much of work from one iteration to the next. In particular, the model changes only slightly and in a very structured way from one iteration to another. The model at iteration $g$ differs from the model at iteration $g-1$ in one of four ways:
\begin{itemize}
    \item {\bf Birth.} One basis function is added to the model at iteration $g$. Thus, we must add a column and row, corresponding to this new basis function, for each $a^{(i)}, b^{(i)}, I_1^{(i)}, I_2^{(i)}, I_3^{(i)}$ matrix. We can then re-compute $\bm C_{ij}$ using \cref{eq:Cijv2}.
    \item {\bf Death.} One basis function has been removed from the model at iteration $g$. If the $m^{th}$ basis function was deleted, then we simply delete the $m^{th}$ row and column from each $a^{(i)}, b^{(i)}, I_1^{(i)}, I_2^{(i)}, I_3^{(i)}$ matrix.
    \item {\bf Mutation.} One basis function has been altered at iteration $g$. If the $m^{th}$ basis function was altered, then we need to recompute the $m^{th}$ row and column of each $a^{(i)}, b^{(i)}, I_1^{(i)}, I_2^{(i)}, I_3^{(i)}$ matrix. 
    \item {\bf Rejection.} No modification was accepted during iteration $g$. The model is the same, except for the regression coefficients $\bm \gamma$ which have been resampled. In this case, we simply use \cref{eq:Cijv2} with the new values for $\bm\gamma$. 
\end{itemize}
Since evaluating integrals is the primary consumer of resources in our estimation procedure, it is clear that there are substantial time-savings to be had. In the case of a birth and a mutation, only $\mathcal O(pM)$ integrals must be evaluated compared to $\mathcal O(pM^2)$ integrals when computing $\bm C$ from scratch. During a death or rejection step, we avoid computing new integrals altogether.

\section{Time and Memory Tradeoff}
 As a brief aside, we also note that when $p$ and $M$ are very large, we can trade memory for time by storing the matrix $I_2^\bullet = \bigodot_{k=1}^p I_2^{(k)}$ and making the approximations
\begin{equation}
\label{eq:I2approx}
\begin{aligned}
    \bigodot_{k\neq i}I_2^{(k)} &\approx I_2^\bullet \oslash \left(I_2^{(i)} + \epsilon\bm J\right) \\[1.5ex]
    \bigodot_{k\not\in \{i,j\}}I_2^{(k)} &\approx I_2^\bullet \oslash \left(I_2^{(i)} + \epsilon\bm J\right) \oslash \left(I_2^{(j)} + \epsilon\bm J\right),
\end{aligned}
\end{equation}
where $\epsilon$ is a small positive number, $\bm J$ is a matrix of ones and $\oslash$ denotes Hadamard division. 

\section{The Effect of Interaction Order on Estimation of C}

Consider a modified simple polynomial function with a three-variable interaction present
$$f_3(\bm x) = x_1^2 + x_1x_2x_3 + x_2^3/9.$$

We simulate data for $n=500$ input variables with $p=10$ and fit a series of BASS models with $J=1, 2, 3$ and $4$. We also fit with a BASS model with the \texttt{birth.type="coinflip"} option, which effectively sets $J=p$. \Cref{tab:Jeffect} gives the out-of-sample RMSPE for each model and the $L_2$ error in estimating $C$ (the true value of $C$ is estimated via Monte Carlo with $1$ million MC samples). We note that the \texttt{concordance} package cannot estimate $C$ in the case where \texttt{maxInt = 1}. 

\begin{table}[!htbp] \centering 
  \caption{RMSPE and $L_2$ error for $C$ for various BASS model fits.} 
  \label{tab:Jeffect} 
\begin{tabular}{@{\extracolsep{5pt}} lccccc} 
\\[-1.8ex]\hline 
\hline \\[-1.8ex] 
& maxInt = 1 & maxInt = 2 & maxInt = 3 & maxInt = 4 & coinflip \\ 
\hline \\[-1.8ex] 
RMSPE & $0.083$ & $0.027$ & $0.002$ & $0.006$ & $0.003$ \\ 
$C$ Error & $-$ & $0.016$ & $0.001$ & $0.003$ & $0.001$ \\ 
\hline \\[-1.8ex] 
\end{tabular} 
\end{table} 

The best results come from the case where $J=3$, which is the true maximum interaction order of the function $f_3$. The second best case is the BASS model fit with the ``coinflipping" birth type, which we propose as a reasonable default when the true interaction order of the model is unknown.

\end{document}